\documentclass[preprint,aps,nofootinbib]{revtex4}
\usepackage{dcolumn}
\usepackage{bm}
\usepackage{epsfig}
\usepackage{graphicx}
\usepackage{amsmath}
\usepackage{amssymb}
\usepackage{mathrsfs}
\usepackage{color}
\usepackage[usenames,dvipsnames]{xcolor}
\usepackage{hyperref}
\usepackage[caption=false]{subfig}
\usepackage[normalem]{ulem}

\def\be{\begin{equation}}
\def\ee{\end{equation}}
\newcommand{\beq}{\begin{equation}}
\newcommand{\eeq}{\end{equation}}
\def\bea{\begin{eqnarray}}
\def\eea{\end{eqnarray}}

\newcommand{\lsim}{ \mathop{}_{\textstyle \sim}^{\textstyle <} }

\newcommand{\GeV}{{\rm{ GeV}}}
\newcommand{\TeV}{{\text{ TeV}}}

\begin{document}

\begin{flushright}
\text{\normalsize MCTP-15-04}\\
\text{\normalsize TUM-HEP 974/15}
\end{flushright}
\vskip 45 pt

\title{Anatomy of Coannihilation with a Scalar Top Partner}
\author{\mbox{\bf A.~Ibarra$^{\;a}$, A.~Pierce$^{\;b}$, N.~R.~Shah${\;^b}$ and S.~Vogl$^{\;c}$}}
\affiliation{
$^a$\mbox{\footnotesize{ Department of Physics, Technische Universtit\"at M\"unchen, Garching, DE-85748}}\\
$^b$\mbox{\footnotesize{Michigan Center for Theoretical Physics, Department of Physics, University of Michigan, Ann Arbor, MI 48109}}\\
$^c$\mbox{\footnotesize{Oskar Klein Centre, Department of Physics, Stockholm University, Stockholm, SE-10961}}\\
\\
\\
}

\begin{abstract}
\vskip 15 pt
We investigate a simplified model of  dark matter where a Majorana fermion $\chi$  coannihilates with a colored scalar top partner $\tilde{t}$.  We explore the cosmological history, with particular emphasis on the most relevant low-energy parameters: the mass splitting between the dark matter and the coannihilator, and the Yukawa coupling $y_\chi$ that connects these fields to the Standard Model top quarks. We also allow a free quartic coupling  $\lambda_h$ between a pair of  Higgs bosons and $\tilde{t}$ pairs. We pay special attention to the case where the values take on those expected where $\tilde{t}$ corresponds to the superpartner of the right-handed top, and $\chi$ is a bino.  Direct detection, indirect detection, and colliders are complementary probes of  this simple model.
\end{abstract}
\thispagestyle{empty}

\maketitle
\newpage

\section{Introduction}

The weakly interacting massive particle (WIMP) is a theoretically attractive candidate for the dark matter, but direct detection bounds are becoming increasingly strong.  Indeed, results from the LUX experiment \cite{Akerib:2013tjd} exclude, depending on the dark matter mass, spin independent dark matter--nucleon scattering cross sections above 10$^{-45}$ cm$^2$.  Such limits represent a challenge to dark matter whose scattering strength is related to annihilation via simple crossing symmetries.  That is, tension arises if the dark matter thermal relic abundance is set by the diagrams that control scattering \cite{Cohen:2011ec,Cheung:2012qy}. For example, this places pressure on so-called ``well-tempered'' scenarios \cite{ArkaniHamed:2006mb} of the minimal supersymmetric Standard Model (MSSM), wherein the proper dark matter abundance is achieved by titrating a non-interacting bino with the appropriate amount of Higgsino, which has full-strength annihilations under the weak interactions.

There are several well-known examples where the crossing symmetry is spoiled \cite{Griest:1990kh}.  One possibility is that a state $Y$ co-inhabits the thermal bath with the dark matter $\chi$ at the time of freeze-out, denoted by the temperature $T_F$.  This is possible if the mass splitting between $\chi$ and $Y$ is not too large, $\Delta M_{\chi Y} \lsim T_{F}$.  In this case, the dark matter abundance is determined not only by the size of the annihilation cross section of $\chi$ to Standard Model~(SM) fields, $\sigma(\chi \chi \rightarrow SM)$, but also by $\sigma(\chi Y \rightarrow SM)$ and, as long as conversions between $\chi$ and $Y$ are sufficiently rapid, $\sigma(YY \rightarrow SM)$.  In the last two {\it coannihilation} cases, the crossing symmetry with direct detection is clearly spoiled:  the interactions of the $Y$ particles with the SM, vitally important for relic density considerations, are irrelevant for direct detection (at tree level).   

Given the current absence of direct detection signals, we are motivated to further consider  the coannihilation scenario, with an eye towards elucidating the physics that determines the relic density.  We consider a simplified model that captures important features of the case in the MSSM where a single light stop squark (say, $\tilde{t}_{R}$) coannihilates with a bino-like neutralino.

Colored particles make for interesting coannihilators since their large QCD cross section somewhat mitigates the otherwise strong mass degeneracy needed for a substantial correction to the relic abundance. 
Furthermore, the LHC clearly calls for a detailed consideration of possible scenarios which include strongly interacting new physics.   For the particular case where the coannihilator is  a top partner, direct detection constraints are less severe than if the coannihilator has large Yukawa coupling with the first two generations, as might be expected for other ``quark partners'' \cite{Garny:2012eb}. In additions, light top partners could provide a testable mechanism for electroweak baryogenesis~\cite{Balazs:2004bu,Balazs:2004ae}.
 Finally,  we note that it might be expected that top partners may be near the bottom of the spectrum of any new TeV scale physics if they are relevant to cutting off the quadratic divergence in the Higgs boson mass squared parameter.

Much of the relevant computations of the underlying physics for stop coannihilation in the full MSSM were already done in the pioneering work of Ref.~\cite{Ellis:2001nx} and were recently reevaluated in~\cite{Ellis:2014ipa}. Often, the MSSM phenomenology is discussed within the context of the Constrained Minimal Supersymmetric Standard Model (CMSSM), and it is not always transparent how to translate results from the $m_0$ -- $m_{1/2}$ plane to the processes that underlie the determination of the relic density.  And yet, there are often a very small number of processes that capture most of the early universe cosmology which can be understood in terms of a few low-energy parameters. This motivates a simplified model approach to the dark matter.  Only the lightest states need be considered for computations of cosmological history, direct detection, and collider signatures. Steps in this direction for a theory of a dark matter accompanied by a colored partner were taken in \cite{Delgado:2012eu,DiFranzo:2013vra,Chang:2013oia,An:2013xka,Bai:2013iqa,Gomez:2014lva,deSimone:2014pda,Garny:2012eb,Garny:2014waa,Kilic:2015vka}.  If desired, a given UV model can be mapped on to this simplified model.  

This article is organized as follows. In Sec.~\ref{s.effstop} we define our model and terminology. We then analyze  the relic density in Sec.~\ref{s.relic}. This includes a basic review of the thermal evolution of a dark matter particle, including coaanihilations. Relevant cross-sections are presented with particular emphasis on their scalings with the parameters of our model. We also detail the effects of the Sommerfeld enhancement. The resulting consistent parameter space is then mapped out, specializing also to the supersymmeteric case. Sec.~\ref{s.Exp} discusses current and future experimental probes of our model, including direct and indirect dark matter detection  as well as collider experiments.  We reserve Sec.~\ref{s.conc} for our conclusions. 
  
\section{Effective stop}\label{s.effstop}

We consider a generalization of the neutralino stop coannihilation scenario. Our ``stop"  is a generic colored scalar with an arbitrary coupling $y_\chi$ to the top quark and the dark matter, which we take to be a gauge singlet Majorana fermion $\chi$.  We consider the scalar in the fundamental representation of the $SU(3)_C$ to be a partner of the right-handed top, so 
\begin{equation}\label{eq.L1}
\mathcal{L}_{\chi}^{int} = y_\chi \bar{t}_R\, \chi \,\tilde{t} + \mbox{h.c.}\; ,
\end{equation}
corresponds to the Lagrangian involving the dark matter field $\chi$ interactions with SM fermions. The MSSM, with $\chi$ identified with the bino, and $\tilde{t}$ identified with the right-handed stop, corresponds to the case $y_\chi^{MSSM} =  \frac{2}{3} \sqrt{2}g_{Y}$, with $g_{Y}$ the hypercharge gauge coupling.  
We also take the quartic coupling between the SM Higgs boson and a pair of stops $\lambda_h$ to be free.   The interaction of $\tilde{t}$ with the SM is given by
\begin{equation}\label{eq.L2}
\mathcal{L}_{\tilde{t}}^{int} =  |D_\mu \tilde{t}|^2 +\lambda_h h^\dagger h \tilde{t}^\dagger \tilde{t}\; .
\end{equation}
In the MSSM without stop mixing, $\lambda_h^{MSSM}=y_t^2=2 m_t^2/v^2\sim1$, where $v=246$ GeV and $y_t$ is the SM Yukawa coupling (we have taken the decoupling limit for the Higgs boson and neglected corrections from supersymmetry breaking). 

In the MSSM, allowing for stop-mixing with large $A$-terms -- as might be motivated by the observed large value of the Higgs boson mass -- can lead to interesting cases with large couplings of (both) stop states to the Higgs boson.  The cosmology is potentially modified in interesting ways with respect to the simpler case presented here, and we plan to explore this in detail in an up-coming publication \cite{Take2}.

It is possible that the scenario defined via Eqs.~(\ref{eq.L1}) and (\ref{eq.L2})   could
be realized in a simple extension of the MSSM by enlarging the gauge sector  and identifying the LSP primarily with the new gaugino. 
  If the corresponding $D$-terms do not decouple, it is also possible to  modify the Higgs boson mass. 
However, our interest is not in model-building, but rather we will use this set-up as an effective parameterization of a model with a small number of degrees of freedom.

\section{Relic Density}\label{s.relic}

The observed relic abundance, $\Omega h^2 = 0.11805 \pm 0.0031$ \cite{PlanckCollaboration2013}, can be easily achieved by thermal freeze-out of weak-scale dark matter, an observation often dubbed the ``WIMP miracle.'' 
The dark matter abundance can be calculated to good accuracy as
\begin{equation}
\label{eqn:OmegaJ}
\Omega h^{2} \approx \frac{8.77 \times 10^{-11} \, \rm{GeV}^{-2}}{g_{\ast}^{1/2}  J(x_{F})}\, ,
\end{equation}
with  $g_{\ast}\sim 80$ is the number of relativistic degrees of freedom and
\begin{equation}
\label{eqn:J}
J(x_{F}) \equiv  \int_{x_{F}}^{\infty} \frac{\langle \sigma {\rm v} \rangle}{x^{2}} dx\, ,
\end{equation}
 where $\sigma$ is the annihilation cross-section, $\rm{v}$ is the relative velocity and the thermally averaged cross section is defined as
\be\label{eqn:thermalsig}
\langle \sigma {\rm v} \rangle =\frac{x^{3/2}}{2 \pi^{1/2}}\int_0^{\infty} ( \sigma {\rm v}) {\rm v}^2 e^{-x {\rm v}^2/4}\, d{\rm v}\, .
\ee
The freeze-out occurring at $x_F\equiv m_{\chi} / T_F$ is determined by the iterative equation
\begin{equation}
\label{eqn:xf}
x_{F} = \log \frac{4.64\times 10^{17}{\rm \, GeV}\, g \; m_\chi \langle \sigma {\rm v} \rangle}{g_{\ast}^{1/2} x_{F}^{1/2}}\, ,
\end{equation}
where $g=2$ for a Majorana fermion.
In the absence of  coannihilations, the thermally averaged annihilation cross section can be expanded as:
\begin{eqnarray}
\langle \sigma_{\chi\chi} {\rm v} \rangle\equiv\langle \sigma ( \chi \chi \rightarrow SM) {\rm v} \rangle = a + b \langle {\rm v}^2 \rangle +\mathcal{O}(\langle {\rm v}^4 \rangle) = a + \frac{6 b }{x} + \mathcal{O}(\frac{1}{x^2})\;.
\end{eqnarray}
yielding 
\begin{eqnarray}
\Omega_{\chi}h^2 \simeq \frac{8.77 \times 10^{-11} \GeV^{-2}  x_F}{\sqrt{g_{\ast(x_F)}} (a +3b/x_F)}\;.
\label{eq:GenOmega}
\end{eqnarray} 
As a rough rule of thumb the freezeout temperature is $T_{F} \sim m_\chi/25$, corresponding to an $x_F  \approx 25$; therefore
\begin{equation}
\Omega_{\chi}h^2 \simeq 0.12 \left(\frac{x_F}{25}\right)
\left(\frac{g_{\ast}}{80}\right)^{-1/2}\left(\frac{a+3b/x_F}{3\times 10^{-26}\mbox{cm}^3/\mbox{s}}\right)^{-1}\;.
\label{eq:GenOmegaNum}
\end{equation}

In the model we consider, the dark matter particles would predominantly annihilate into a pair of tops, $(\chi \chi \rightarrow t \bar{t})$, via the $t$ and $u$-channel exchange of a $\tilde{t}$. The $s$-wave contribution to this is given by:
 \be \label{eq.chitt-a}
a=  \frac{3 \, m_t^2 \, y_{\chi }^4 \sqrt{m_{\chi }^2-m_t^2}}{32\, \pi \,  m_{\chi }
   \left(m_{\tilde{t}}^2-m_t^2+m_{\chi }^2\right){}^2}\,.
   \ee
   As a chirality flip of the top in the final state is required,  $a$ is  proportional to
    $m_t^2$, and therefore in the limit where $\chi$ is much heavier than the top, the dominant contribution is due to the velocity suppressed $p$-wave contribution:
 \be
 \label{eq.chitt-b}
b \simeq \frac{m_{\chi }^2 y_{\chi }^4 \left(m_{\tilde{t}}^4+m_{\chi }^4\right)}{16 \pi 
   \left(m_{\tilde{t}}^2+m_{\chi }^2\right){}^4}\;.
 \ee
We present only the $m_t \rightarrow 0$ limit of $b$ here, however the full mass dependence is always used in our numerical results, which are obtained by solving the relevant Boltzman equation(s) numerically with the help of {\tt MicrOmegas 3.3}~\cite{Belanger:2013oya}.

As we will review below, the above expressions must be modified if additional degrees of freedom, e.g. $\tilde{t}$, are still present in the thermal bath during freeze-out.  
In addition, it is well known that the annihilation cross section of charged non-relativistic particles can be modified by non-perturbative corrections: the Sommerfeld effect~\cite{Orig:Somm,Hisano:2003ec,Hisano:2004ds,ArkaniHamed:2008qn}. As $\tilde{t}$ is charged under QCD, these corrections can become important and lead to significant modifications of the relic density~\cite{deSimone:2014pda,Baer:1998pg,Freitas:2007sa,Hryczuk:2011tq}.   Finally, we caution the reader that higher order corrections to stop coannihilations can be significant~\cite{Harz:2014tma,Harz:2014gaa}, and the accuracy of the relic density calculation lags behind the precision of the observations by Planck.

\subsection{Coannihilation}

The canonical calculation of the relic density needs to be modified if other particles of the dark sector are close in mass with the dark matter~\cite{Griest:1990kh}. Not only dark matter annihilations, but also processes involving the next to lightest particles in the dark sector can contribute to an effective annihilation rate.
Quantitatively, this corresponds to replacing the annihilation cross section $\sigma_{\chi \chi} {\rm v}$ in the Boltzmann equation with an effective cross section given by~\cite{Griest:1990kh}\begin{eqnarray}
\sigma_{eff} {\rm v} = \sum_{i,j} \frac{n_i^{eq} n_j^{eq}}{(\sum_k n_k^{eq})^2}\sigma_{ij} {\rm v}\;,
\end{eqnarray}
where $n_i^{eq} = g_i (m_i T/ (2 \pi))^{3/2} e^{-m_i/T}$; $m_i$ is the mass of the particle $i$, and $g_i$ counts the number of internal degrees of freedom. 
The relic abundance can be approximately found using Eqs.~(\ref{eqn:OmegaJ})-(\ref{eqn:xf}), now utilizing this effective cross section.  We emphasize, however,  that our numerical results rely on numerical solution to the Boltzman equation via {\tt MicrOmegas 3.3}~\cite{Belanger:2013oya}.

In the model studied here, coannihilation of $\chi$ with $\tilde{t}$ as well as the annihilation of $\tilde{t} \tilde{t}^*$ pairs are relevant for the calculation of $\sigma_{eff} {\rm v}$~\cite{Ellis:2001nx}.
In general, the contribution due to processes with $\tilde{t}$ in the initial state require a convolution of  the annihilation cross section and its thermal abundance.   The abundance of the heavier state is suppressed relative to that of $\chi$ by additional factors of $e^{-\Delta m /T_F}$, where $\Delta m = m_{\tilde{t}} -m_{\chi}$. Due to this exponential dependence, the relic density is extremely sensitive to the mass splitting between $m_\chi$ and $m_{\tilde{t}}$. 
Contributions to the relic density from the coannihilations of a pair of ${\tilde{t}}$s are doubly exponentially suppressed compared to $\chi\chi$ annihilations, and the annihilations of  $\chi$ with a ${\tilde{t}}$ are singly exponentially suppressed.  For $m_{\tilde{t}} \gtrsim 1.2\, m_{\chi}$, coannihilations can safely be neglected. Note that the possibility of coannihilations provides a lower limit on $\Delta m$ as a function of $m_{\chi}$.   If the mass splitting is too small, then coannihilations will be too effective, owing to the large irreducible (QCD) cross section for $\tilde{t} \tilde{t^{\ast}}$ annihilations:
  \be
 \sigma {\rm v} (\tilde{t} \tilde{t}^*\rightarrow gg)= \frac{7 g_s^4}{216 \, \pi \,  m_{\tilde{t}}^2} \, .
 \ee 
 
 \begin{table}[!pt]
 \begin{tabular}{ | c |   m{8 cm} |  }
 \hline
 \hline
 Channel & \hskip 1 cm $ \sigma {\rm v}$ \\
 \hline
 \hline
 &\\
$\chi \chi \rightarrow t \bar{t} $ & \hskip 1cm $ \frac{3 \, y_{\chi }^4 \, m_t^2 }{32\, \pi \,(r^2+1)^2  m_{\chi }^4}$ \\
   &\\
   \hline
 &\\
 $\chi \tilde{t} \rightarrow g t$ & \hskip 1cm $  \frac{g_s^2 y_{\chi }^2}{24 \,  \pi \,  r (r+1)\, m_{\chi }^2} \left[1-\frac{m_t^2}{m_{\chi }^2\, (r+1)^2 }\right]$\\
 &\\
$\chi \tilde{t} \rightarrow h t $ & \hskip 1cm $\frac{y_\chi^2 m_t^2}{64 \pi v^2} \frac{1}{ r (1+r) m_\chi^2}\left[1+  \frac{\lambda_h v^2}{r^2 m_\chi^2}\left(\frac{v^2}{m_t^2}\lambda_h +\frac{6 r}{1+r} \right)  \right] $ \\
&\\
$\chi \tilde{t} \rightarrow Z t $ & \hskip 1cm $\frac{y_\chi^2 m_t^2}{64 \pi v^2} \frac{1}{ r (1+r) m_\chi^2}\left[1-\frac{m_t^2 \,r}{m_{\chi }^2\, (r+1)^2 }\right] $ \\
&\\
 $ \chi \tilde{t} \rightarrow W b$ & \hskip 1cm $\frac{y_{\chi }^2 m_t^2}{32 \pi v^2} \frac{1}{ r (r+1) m_\chi^2}   \left[1+\frac{2 m_W^2}{(r+1)^2 m_{\chi }^2}\right] $  \\
&\\
\hline
&\\
$\tilde{t} \tilde{t}^*\rightarrow gg$ &\hskip 1cm  $\frac{7 g_s^4}{216 \, \pi \,  m_{\tilde{t}}^2}$ \\
&\\
$  \tilde{t} \tilde{t}^*\rightarrow h h $ & \hskip 1cm $ \frac{\lambda_h^2}{192\, \pi\, m_{\tilde{t}}^2 }\left[1+\frac{v^2}{m_{\tilde{t}}^2 }\left(\lambda_h +\frac{3}{4} \frac{m_h^2}{v^2}\right)  \right]^2$  \\
&\\
  $\tilde{t} \tilde{t}^*\rightarrow ZZ$ & \hskip 1cm $ \frac{\lambda_h^2}{192\, \pi\, m_{\tilde{t}}^2 }  \left(1-\frac{m_Z^2}{m_{\tilde{t}}^2}\right)$  \\
  &\\
 $ \tilde{t} \tilde{t}^*\rightarrow W^+W^-$ & \hskip 1cm $ \frac{ \lambda _h^2 }{96\, \pi \,  m_{\tilde{t}}^2} \left[ 1- \frac{(m_h^2+2 m_W^2)}{2 m_{\tilde{t}}^2} \right]$  \\
 &\\
 \hline
\end{tabular}
\caption{Dominant contributions to the cross-sections relevant for setting the relic density, when $m_\chi$ and $m_{\tilde{t}}$ are much larger than the weak scale. The mass splitting is parametrized by the ratio $r=m_{\tilde{t}}/m_\chi$.  
}\label{T.XS}
\end{table}

In addition to $(\tilde{t} \tilde{t}^*\rightarrow g g)$, channels involving electroweak bosons can be relevant, specially in the large $\lambda_h$ regime.  Because a relatively small number of processes contribute to the determination of the relic density in our model, we find it instructive to reproduce simple expressions for the relevant cross sections, tabulated in Table \ref{T.XS}. Only the dominant contributions in the limit $m_\chi$ and $m_{\tilde{t}}$ much larger than the weak scale, are listed. We have separated out the  cross-sections in Table \ref{T.XS} into three parts, depending on the initial state. The relevance of each initial state depends on the  mass splitting between $\chi$ and $\tilde{t}$, which we parameterize by the ratio of their masses, $r=m_{\tilde{t}}/m_\chi$. All cross-sections were computed by implementing our Simplified Model in {\tt CalcHEP 3.4}~\cite{Belyaev:2012qa}. 

As  seen in Table \ref{T.XS},  the channels scale differently with the dark sector couplings $y_\chi$ and $\lambda_h$. Cross sections of the processes initiated by  $\chi \tilde{t}$, which are thermally less suppressed than $\tilde{t} \tilde{t}^\ast$, are proportional to $y_\chi^2$ and therefore depend rather strongly on the dynamics of the new sector. Typically, the dominant channel here is $(\chi \tilde{t} \rightarrow g t)$, which scales as $y_\chi^2 g_s^2$, but, 
final states with the Higgs or electroweak gauge bosons are not irrelevant and the $ht$ final state can receive a significant enhancement if $\lambda_h > 1$. On the other hand, $(\tilde{t} \tilde{t}^* \rightarrow gg)$, typically the most relevant process for the most degenerate cases, 
depends only on the strong coupling and $m_{\tilde{t}}$ and does not 
depend on $y_\chi$ or $\lambda_h$. Finally, the leading contribution to the annihilation of $\tilde{t} \tilde{t}^*$ pairs into pairs of massive electroweak
bosons are controlled by the interactions of $\tilde{t}$ with the Higgs boson and therefore the annihilation rates scale as $\lambda_h^2$. Typically these processes are subdominant compared to the annihilations into gluons, however, they can become more relevant once $\lambda_h > 1$ is considered.

\subsection{QCD Sommerfeld Effect}\label{s:SE}

The annihilation of two scalar top partners, which plays a pivotal role in the determination of the dark matter relic density, is affected by the non-perturbative Sommerfeld effect. As a result, the annihilation cross-sections at the time of freeze-out in the different final states  can significantly differ from the Born-level values given in Table \ref{T.XS}. We briefly review here the formalism to include the Sommerfeld enhancement in the calculation, following closely the approach of Ref.~\cite{deSimone:2014pda}.  

We consider an annihilation process where the particles in the initial state interact with each other via a long-range interaction described by a central potential. For a cross section with  partial wave expansion $\sigma = \sum_{l=0} a_l {\rm v}^{2l -1}$, the Sommerfeld corrected cross-section is given by
\beq \label{eq:sigmaS}
\sigma^S =\sum_{l=0} S_l \; a_l {\rm v}^{2 l -1},
\eeq 
namely, the non-perturbative effects can be calculated separately for each partial wave and are encoded in the enhancement factor $S_l$. 

 Interactions mediated by a massless particle generate a Coulomb-like potential of the form $V=\alpha /r$, attractive for $\alpha<0$, where $\alpha$ is the potential strength. The enhancement factor due to such a potential for the $s$-wave component of the annihilation cross section is given by
\beq \label{eq:S0}
S_0=\frac{-\pi \alpha/\beta}{1 - e^{\pi \alpha/\beta}}\;,
\eeq
 while for  $l>0$~\cite{Cassel:2009wt,Iengo:2009ni}:
\beq\label{eq:Sl}
S_{l>0}= S_0\times \prod_{n=1}^l\left(1+\frac{\alpha^2}{4 \beta^2n^2}\right)\, ,
\eeq
where $\beta={\rm v}/2$~(recall that {\rm v}  is the relative velocity of the incoming particles).

It follows from this expression that when ${\rm v} \ll 1$, $S_{l>0}\sim S_0 {\rm v}^{-2l}$, which seems to jeopardize the convergence of the partial wave expansion, Eq.~(\ref{eq:sigmaS}). However, we will be interested in potentials where $\alpha \sim 0.1$, therefore the terms with high $l$ are suppressed by a factor $(0.1^{2l}/l!^2)$ which ensures the convergence of the series. This implies that, in practice, it suffices to keep only the first few terms in the expansion. Then, casting the Born-level cross section as
\begin{eqnarray}
\sigma &=& \frac{a_0}{{\rm v}}+a_1\; {\rm v}  + \delta_l  \;,
\end{eqnarray}
where $\delta_l$ includes the higher $l$ contributions to the cross-section, the Sommerfeld corrected cross-section is well approximated by 
\beq
\sigma^S \simeq S_0\frac{a_0}{{\rm v} }+ S_1a_1\; {\rm v} + S_2\delta_l\, ,
\eeq
which we implement in our numerical work.

The annihilations  $(\tilde{t}\tilde{t}^*\rightarrow hh, V_1 V_2)$, with $V_1$, $V_2$ being gauge bosons, can be significantly affected due to the potential generated by gluon exchange. 
Prior to the QCD phase transition, this potential can be approximately described by a Coulomb-like potential \cite{Fischler:1977yf,Strumia:2008cf}
\begin{equation}
V(r) \approx C \frac{\alpha_s}{r} = \frac{\alpha_s(\mu = 1/r)}{2 r } [C_Q -C_R- C_{R'}]\;,
\label{eq:QCDpotential}
\end{equation}
where $\alpha_s$ is the strong coupling constant evaluated at the scale $\mu = 1/r$, $C_Q$ is the quadratic Casimir coefficient of the color representation of the final state and $C_R$ and $C_{R'}$ are the quadratic Casimir coefficients of the incoming particles. In the case of scalar top partners, $C_R=C_{R'}=C_3=C_{\bar{3}}=4/3$, whereas  $C_Q$ depends on the annihilation channel. The relevant final states are characterized  by $C_1=0$ for a singlet and $C_8=3$ for an octet representation. As the sign of $C$ depends on $C_Q$, the potential can be both attractive or repulsive, depending on the color representation of the final state.  Consequently, annihilations into color singlet final states, e.g., $hh$, receive a universal enhancement,  whereas the case is more complicated for annihilations into final state which can be in more than one color representation.

An initial $\tilde{t}\tilde{t}^\ast$ state is decomposed as $\boldsymbol{3}\otimes \boldsymbol{\bar{3}}= \boldsymbol{1}\oplus \boldsymbol{8}$.\footnote{A $\tilde{t}\tilde{t}$ initial state, which decomposes as a $\boldsymbol{\bar{3}} \oplus \boldsymbol{6}$ can be relevant for larger masses than those considered here.  We include it, and the relevant Sommerfeld enhancement, in our numerics.} The QCD potential corresponding to these states is 
\beq
V=\frac{\alpha_s}{r}\times \left\{\begin{array}{c c}-4/3 & \mbox{ for the } \boldsymbol{1}, \\  +1/6 & \mbox{ for the } \boldsymbol{8}. \end{array}\right.
\eeq
The most relevant final states for cosmology resulting from such an initial state are $gg$, $hh$, $W^+W^-$ and $ZZ$. For completeness we also include the $\gamma g$ and $Z g$ final states. Only the $gg$ final state can have both the singlet and octet states, the others are either only color singlets or color octets.   For the $gg$ final state the  Sommerfeld enhancement factor for each partial wave reads \cite{deSimone:2014pda}
\beq\label{eq:Stogg}
S_l^{\tilde{t}\tilde{t}^*\rightarrow gg}=\frac{2}{7}\; S_l\Big|_{\alpha= -\frac{4}{3}\alpha_s}+ \frac{5}{7}\; S_l\Big|_{\alpha=\frac{1}{6}\alpha_s},
\eeq
where the coefficients can be derived from a simple group theory calculation \cite{deSimone:2014pda}.
Whereas for the singlet/octet only final states, 
\beq\label{eq:StohhVVtt}
S_l^{\tilde{t}\tilde{t}^*\rightarrow hh, ZZ,W^+W^-}=S_l\Big|_{\alpha= -\frac{4}{3}\alpha_s}\;, \qquad \qquad S_l^{\tilde{t}\tilde{t}^*\rightarrow Zg, \gamma g}=S_l\Big|_{\alpha= \frac{1}{6}\alpha_s}\;,
\eeq
with $S_l$ as given in Eqs.~(\ref{eq:S0}) and (\ref{eq:Sl}).

The above description of the Sommerfeld effect holds if a definite QCD representation can be assigned to the initial state. It has been argued in Refs.~\cite{Berger:2008ti,Baer:1998pg} that this assumption might not be correct in the thermal bath present at freeze-out:  rapid interactions with gluons could continuously  change the color state of the initial particles and thus prevent the formation of a definite QCD color state. In this case, a color averaged initial state could be considered instead. Qualitatively, this prescription corresponds to the replacement of the individual QCD potentials of different representations in Eq.~(\ref{eq:QCDpotential}) with an averaged potential.  In that case all the cross section with the same initial state should be scaled by a single Sommerfeld factor independently of the final state.   Performing the color average, we find the following QCD potential 
\beq
V=\frac{\alpha_s}{r}\times \left( -\frac{11}{42}\right) .\eeq
We will refer to the above as the \emph{color averaged Sommerfeld} effect.   We explore the effect of these two prescriptions in what follows, but primarily use the ``color coherent" prescription.  Regardless of the prescription chosen, the picture remains qualitatively the same.  A definitive answer to the proper treatment of these thermal effects lies beyond the scope of this work.  
An additional thermal effect arises from the screening of the QCD potential due to the gluon's plasma mass.
As was shown in Ref.~\cite{deSimone:2014pda}, including the thermal mass of the gluon has an imperceptible impact on the relic density. 

Finally, it should be kept in mind  that dark matter annihilation and the Sommerfeld effect are processes happening at different scales. Consequently, it is not appropriate to evaluate the strong coupling $\alpha_s(\mu)$ which enters the enhancement factors $S_l$ at the energy scale of the hard annihilation process $\mu \approx 2m_\chi $.  Rather, the Sommerfeld scale $\mu \approx p$, where $p$ is the momentum of the annihilating particles, should be used. 
This could bring into question the validity of our calculation if $\alpha_s(\mu)$ enters the non-perturbative regime. However, this is not an issue because the low  abundance of such very small momentum particles  renders  the details of the running of $\alpha_s$
at these scales unimportant.

\subsection{Numerical Results}\label{Relic.Numb}

We implemented our simplified model using {\tt CalcHEP 3.4}~\cite{Belyaev:2012qa} in {\tt MicrOmegas 3.3}~\cite{Belanger:2013oya} and incorporated the appropriate 
Sommerfeld factors for the different annihilation channels relevant for coannihilation of a scalar top partner as described above. We now discuss the cosmology of this model.
Experimental constraints and future prospects on the presented parameters due to collider and other astrophysical observations will be discussed in Sec.~\ref{s.Exp}. 

\subsubsection{MSSM:  $\boldsymbol{\tilde{B}+\tilde{t}_R}$ }\label{s:SUSY}

\begin{figure}[tb]
\subfloat[]{\includegraphics[width=3.2in, angle=0]{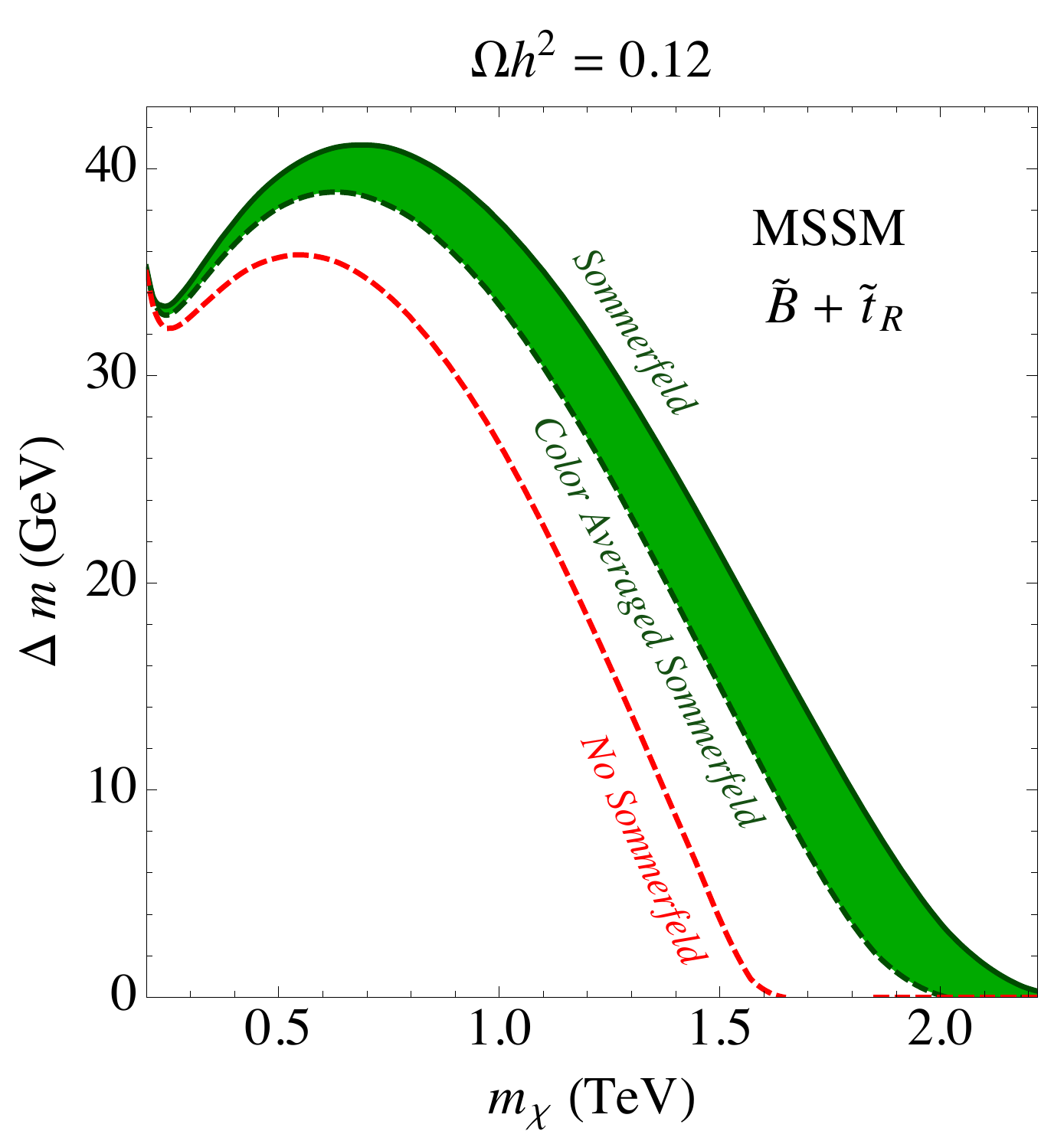}\label{f:SUSYDelta}}~~
\subfloat[]{\includegraphics[width=3.2in, angle=0]{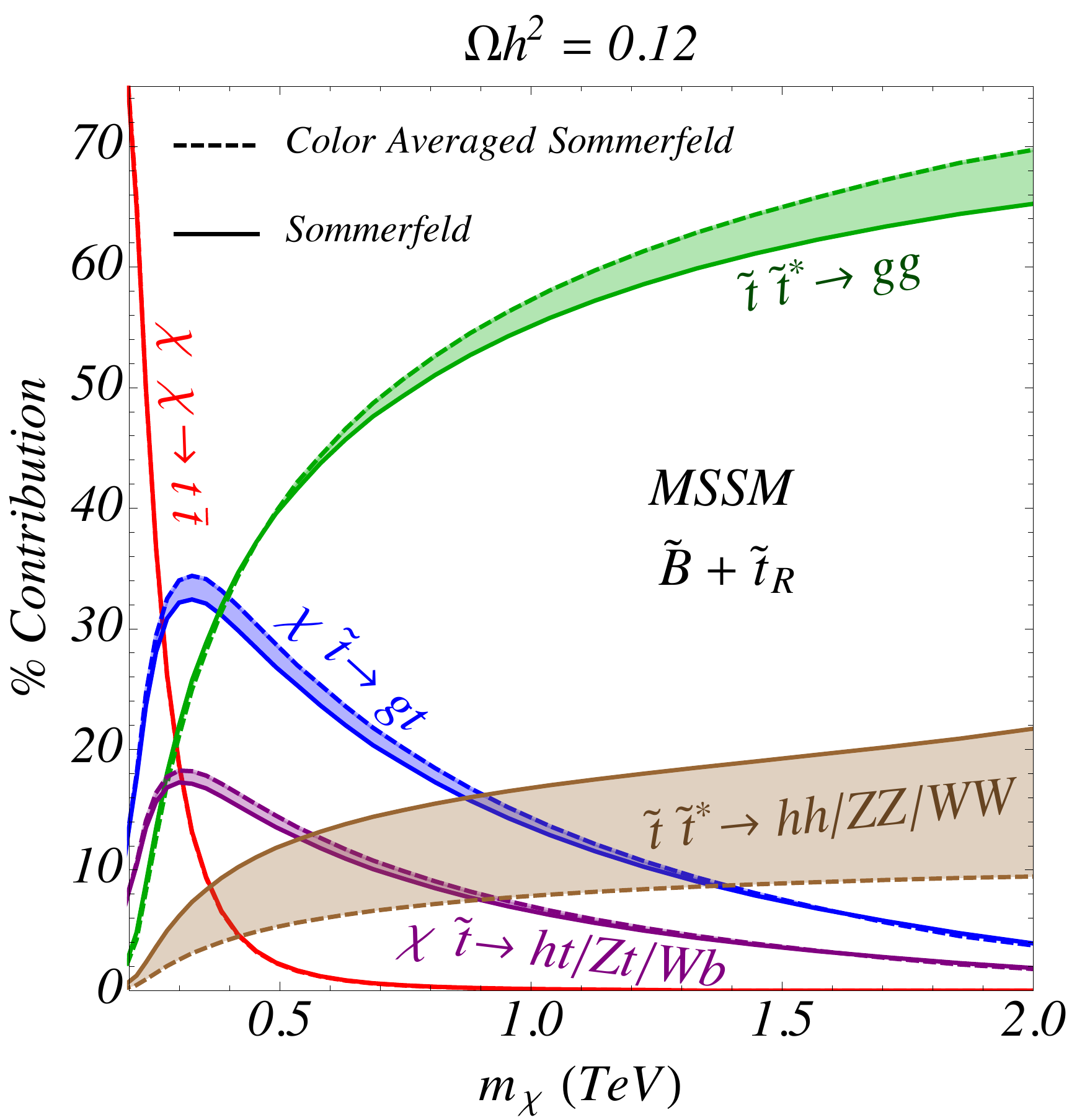}\label{f:SUSYCont}}
\caption{\label{fig:SUSY}{\em  a) The mass difference, $\Delta m$, required to obtain the correct relic density with (green) and without (dashed red) the Sommerfeld effect. The green band spans the mass region between the color averaged (dashed) and the color summed (solid) calculation of the Sommerfeld enhancement. \\ b) The percent contribution of the various annihilation channels to the relic density,   varying $\Delta m$ as needed for each $m_{\chi}$ to satisfy $\Omega h^2 = 0.12$.}}
\end{figure}

The MSSM limit of our simplified model is reached by identifying $\chi$ with a bino like neutralino $\tilde{B}$ and the scalar $\tilde{t}$ is specifically a right handed stop $\tilde{t}_R$. The second stop is heavy and therefore decoupled. In terms of the MSSM parameters, the demand that the stop mass eigenstates be the gauge eigenstates, explicitly requires a vanishing mixing parameter in the stop sector, $X_t = 0$. For consistency with the observed Higgs boson mass, the heavy stop would then need to be PeV scale. The couplings in this case approximately reduce to: $y_{\chi}^{MSSM} = \frac{2}{3} \sqrt{2}g_{Y}$ and $\lambda_h=y_t^2$.

In Fig.~\ref{f:SUSYDelta} we show the mass splitting required between the $\tilde{B}$ and the $\tilde{t}_R$ to accommodate the observed relic density. The red dashed curve denotes values neglecting the Sommerfeld effect. The band of green values is obtained by considering the two prescriptions for the Sommerfeld enhancement factor, as detailed in Sec.~\ref{s:SE}. 

Fig.~\ref{f:SUSYCont} shows the annihilation channels giving the dominant contributions to the relic density as a function of the $\tilde{B}$ mass. We have not displayed $(\tilde{t} \tilde{t}^* \to Zg / \gamma g )$, which combine to contribute $\lsim 10 \%$. The importance of the different channels depends strongly on $m_\chi$.  The channel $(\tilde{t} \tilde{t}^* \to gg)$ is only  dominant when $m_\chi$ is greater than about 700 GeV. If the relic density is computed just using this channel, one would obtain a mass difference approximately 10 GeV smaller than the results shown here~\cite{Delgado:2012eu,deSimone:2014pda}.   Again the width of the band corresponds to the different treatment of the Sommerfeld prescriptions.  For the remainder of the paper, we will utilize the color coherent prescription, bearing this uncertainty in mind.

\subsubsection{Simplified Model}

In our simplified effective stop model, the couplings $y_\chi$ and $\lambda_h$ are free parameters. The quartic coupling $\lambda_h$ only impacts the coannihilation channels which are related to the Higgs boson, either in the intermediate or the final state.

\begin{figure}[tb]
\subfloat[]{\includegraphics[width=3.2in, angle=0]{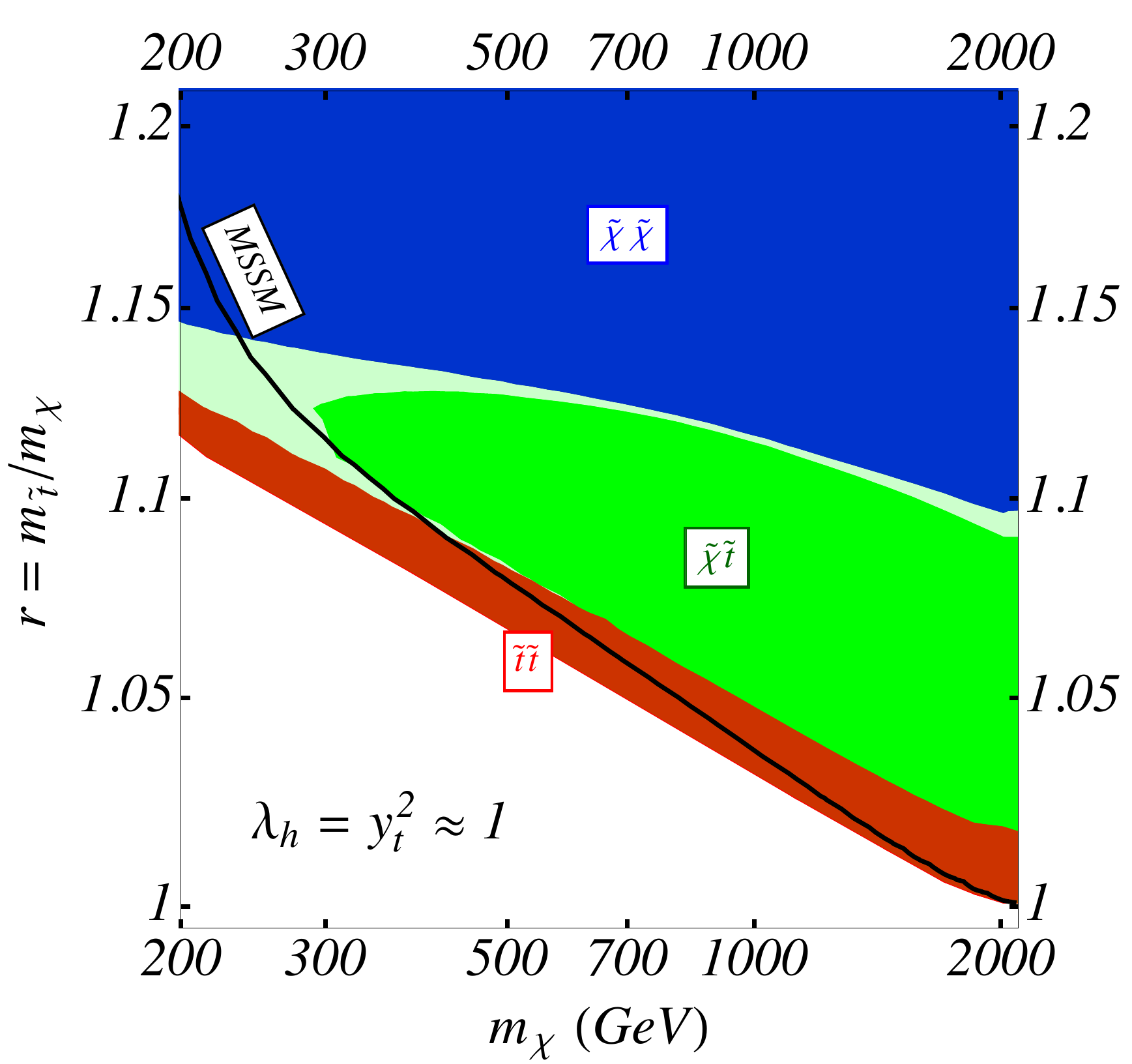}\label{fig:InStatesl1}}~~
\subfloat[]{\includegraphics[width=3.2in, angle=0]{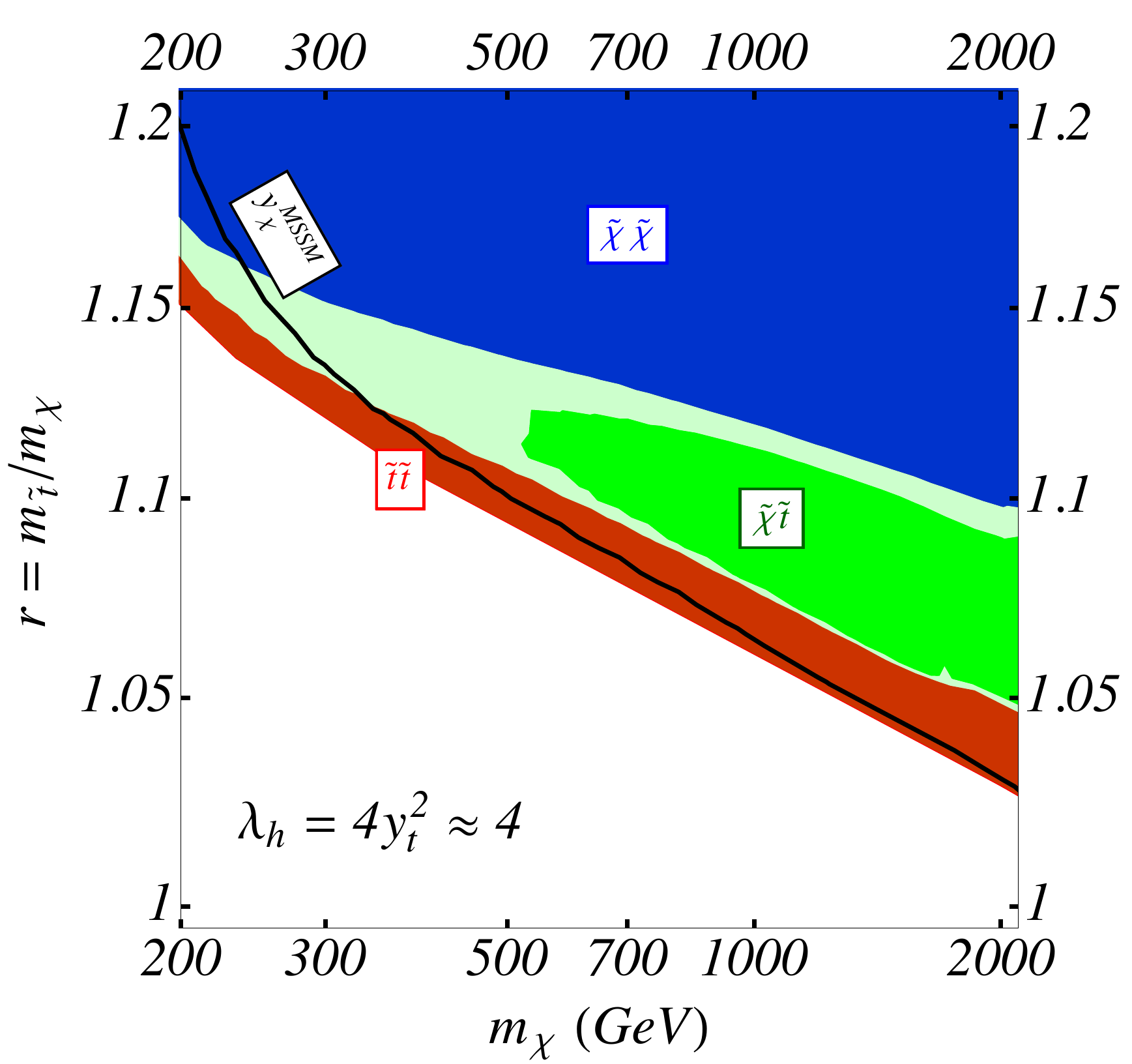}\label{fig:InStatesl4}}
\caption{{\em  Regions where the relic density is dominated by different initial states. a) $\lambda_h=y_t^2\approx 1$. The black line correspond to the minimal MSSM $\tilde{B} + \tilde{t}_R$ model.  b)  $\lambda_h= 4 y_t^2\approx 4$. The black line corresponds to $y_{\chi}^{MSSM} = \frac{2}{3} \sqrt{2}g_{Y}$.}}
\end{figure}

As a means of analyzing this multidimensional parameter space we will first fix $\lambda_h$ to a given value.  
Then, we find the minimum $r=r_{min}$ that can yield the correct abundance, found by saturating the necessary annihilation cross section by the  $gg$ final state alone.  For all values of $r> r_{min}$ there exists a $y_\chi$ such that the experimentally observed relic density can be found.  In Fig.~\ref{fig:InStatesl1} and \ref{fig:InStatesl4} we show the contributions to the relic density from different initial states in the $m_\chi$ -- $r$ plane for $\lambda_h=y_t^2 \approx 1$~(left) and $\lambda_h  \approx 4 y_t^2\approx4$~(right) respectively, such that the relic density is fixed to be $\Omega h^2=0.12$ by varying $y_\chi$. The white portion of the plots denotes the region where the experimentally observed relic density cannot be obtained thermally, even with $y_\chi=0$. The red, bright green and blue regions denote where the dominant initial states~(i.e. contribute more than 50\%) are $\tilde{t}\tilde{t}$, $\chi\tilde{t}$ and $\chi\chi$ respectively. In the light green shaded region several of the three initial states are of similar strength and no dominant channel can be identified. The black line in both plots shows where $y_\chi$ is equal to the MSSM value of $2\sqrt{2}\,g_Y/3$. Comparing the left and right panels we can see the impact of changing the value of the quartic coupling $\lambda_h$ on cosmology. Defining coannhilation as the region where the $\chi\chi$ initial state is not the dominant contribution, we can see that, apart from the very low mass region, the value of $r$ for the cross-over from coannihilation to annihilation (i.e. the transition to the blue region) for a given $m_\chi$, is relatively insensitive to $\lambda_h$.  Large $\lambda_h$ does effect the coannihilation region: For $\lambda_h \approx 4$ the red region  shifts approximately 3\% upwards in $r$ across the $m_\chi$ range considered with respect to the $\lambda_h \approx 1$ case. At very low masses, there is a change in this behavior.   This can be understood by 
 comparing the $\lambda_h$  dependence of the three channels, $\tilde{t} \tilde{t}^{\ast} \rightarrow$ $hh$, $WW$, $ZZ$:  At lower masses of $\tilde{t}$ the scaling of the $hh$ channel  transitions from $\lambda_h^2$ to $\lambda_h^4$ ~(cf. Table \ref{T.XS}). In any event, as we will see, a light scalar top partner with such a large quartic coupling to Higgs bosons would cause significant deviation in gluon fusion, which is most certainly excluded by Higgs coupling measurements~\cite{Khachatryan:1979247}.

\begin{figure}[tb]
{\includegraphics[width=4.2in, angle=0]{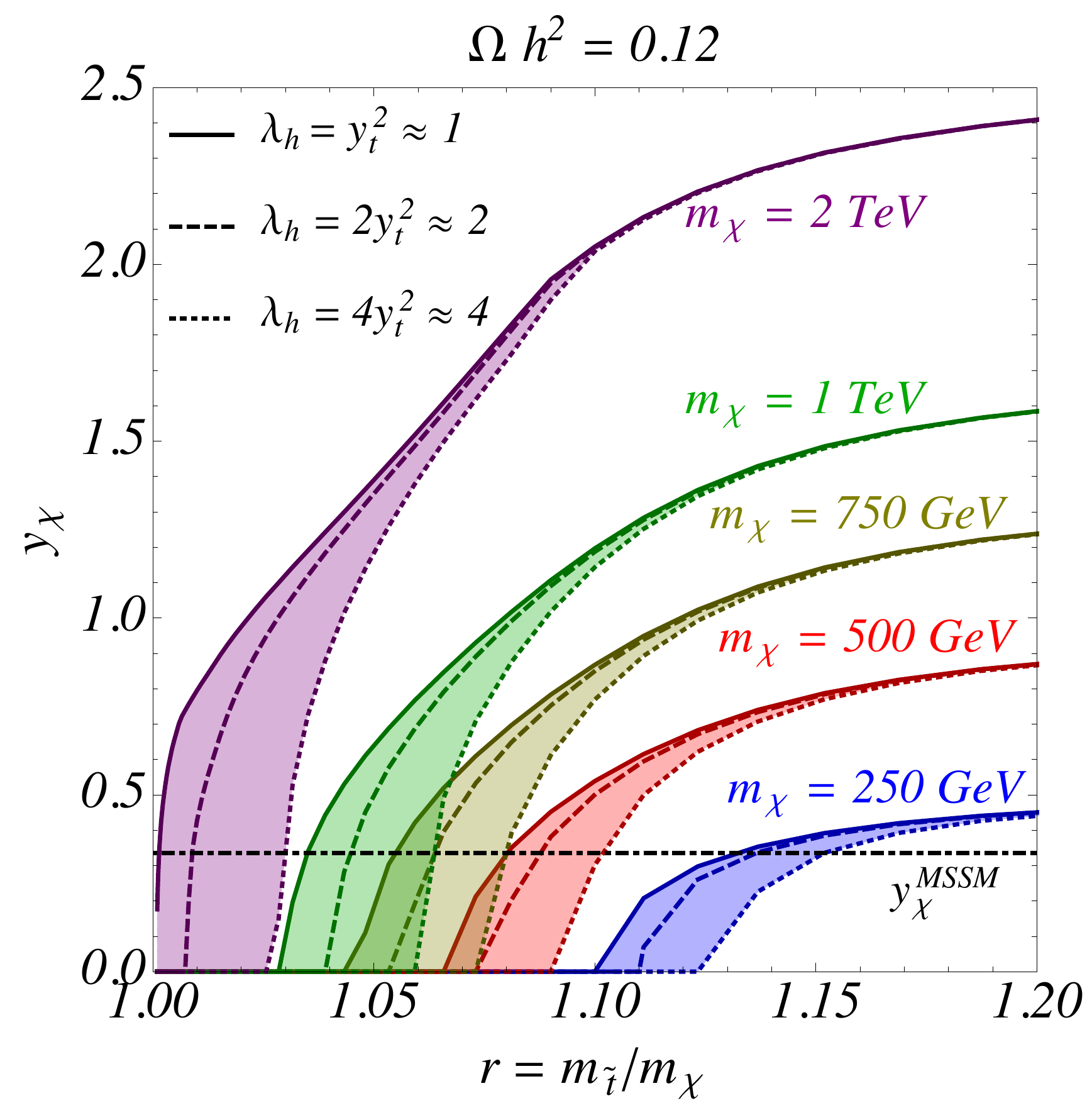}}
\caption{\label{fig:gBRR}{\em  $y _\chi$ as a function of $r = m_{\tilde{t}} / m_\chi$ for various choices of the dark matter mass. The shaded regions corresponding to each mass are bounded by $\lambda_h=  y_t^2 \approx 1$ (solid) and $\lambda_h = 4 y_t^2\approx 4$ (dotted). The $\lambda_h= 2 y_t^2 \approx 2$ value is denoted by the dashed lines in each shaded region.  The horizontal black dot-dashed line denotes the value of the MSSM coupling, $y_\chi^{MSSM}$.}}
\end{figure}

Fig.~\ref{fig:gBRR} shows values of $y_\chi$ required as a function of $r$ for different $m_\chi$ to obtain $\Omega h^2=0.12$. The width of the colored bands captures the effect of varying $\lambda_h$, with the solid~(dotted) lines corresponding to $\lambda_h \approx 1\;(4)$. The $\lambda_h \approx 2$ value is shown by the dashed line in the middle of the shaded regions. The horizontal black dot-dashed line denotes $y_{\chi} =y_\chi^{MSSM}= \frac{2}{3} \sqrt{2}g_{Y}$, corresponding to the MSSM value. The values of $r$ where the $y_{\chi}^{MSSM}$ line intersects the $\lambda_h \approx 1$ lines for each $m_\chi$ corresponds to the mass splittings shown as the solid green curve in Fig.~\ref{f:SUSYDelta}. The narrowing of the colored bands for increasing $r$ shows the transition from coannihilation into the self-annihilation region and confirms that the value of $\lambda_h$ is only relevant for cosmology when there is a significant contribution to the relic density from coannihilation. Further, as also seen from the previous figures,  even extreme values for $\lambda_h$ only result in a shift of at most 3\% for the needed $r$ (for fixed $m_{\chi}$) throughout the mass range we consider.  In contrast, values of $y_\chi$ comfortably below perturbativity constraints can allow for large mass splittings, easily allowing for $r\gtrsim 1.2$ even for the heaviest $m_\chi$ considered.

\begin{figure}[tb]
{
\includegraphics[width=4.2in, angle=0]{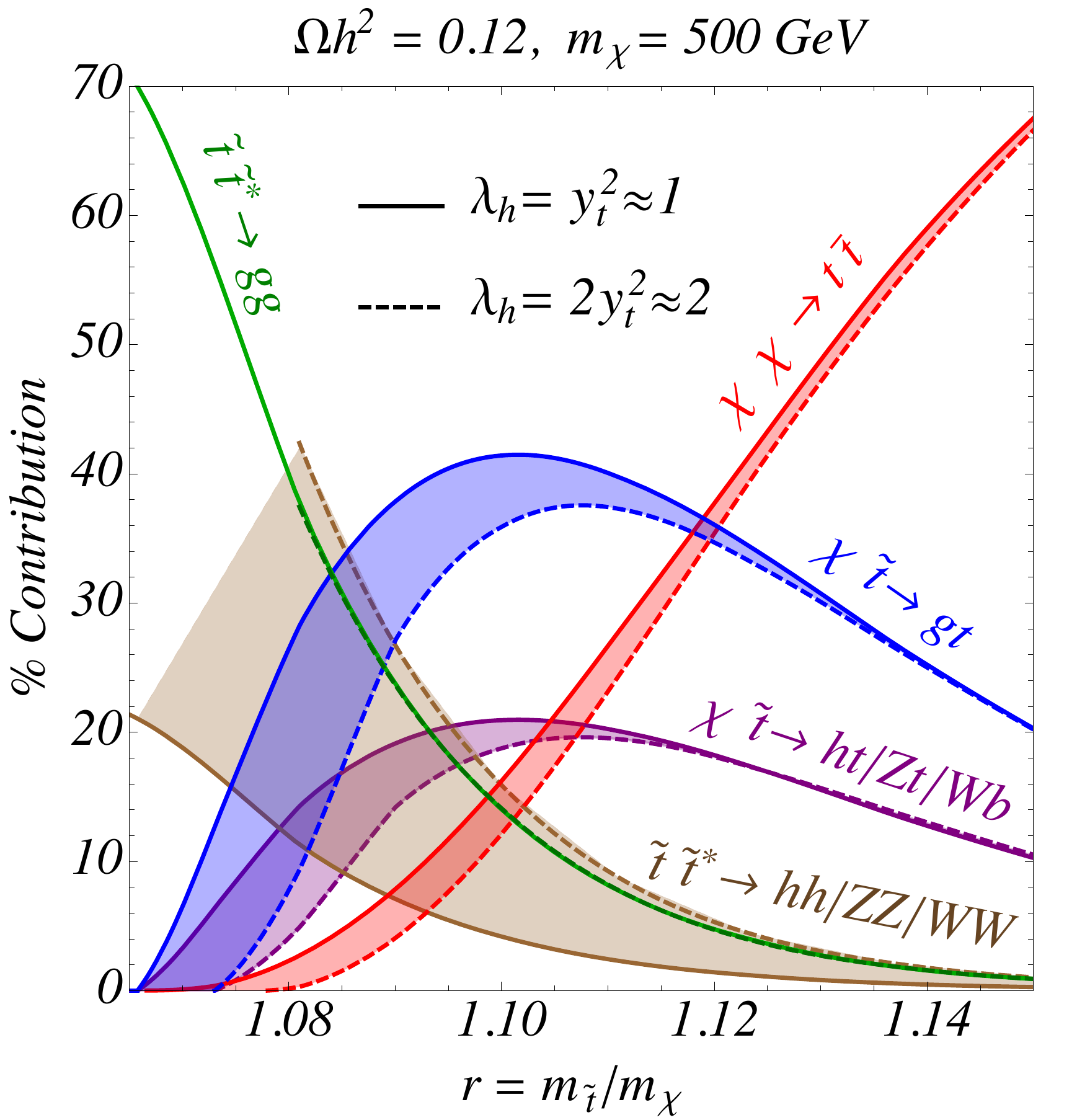}}
\caption{\label{fig:PercCont}{\em Percentage contributions from relevant annihilation channels as a function of $r=m_{\tilde{t}}/m_\chi$, resulting in $\Omega h^2=0.12$ for $m_\chi=500 \;\rm GeV$ . The shaded bands denote the variation in  $\lambda_h$ from 1 to 2. The corresponding $y_\chi$ is as shown in Fig.~\ref{fig:gBRR}}. 
}
\end{figure}

In Fig.~\ref{fig:PercCont} we show the relative percentage contributions from different channels to the total relic density, given $\Omega h^2=0.12$, as a function of $r$ for $m_\chi=500$ GeV. The bands correspond to the variation of $\lambda_h$ between 1 (solid lines) and 2 (dashed lines).  
The values of $y_\chi$ correspond to those shown in Fig.~\ref{fig:gBRR}. As $\lambda_h$ increases, the smallest value of $r$ consistent with the proper relic density increases.  Note that since the contribution due to the $(\tilde{t} \tilde{t}^* \to gg)$ channel depends only on the mass splitting for a given $m_\chi$, it doesn't get perturbed by the change in $\lambda_h$. Therefore, the increase in  $(\tilde{t} \tilde{t}^* \to hh/ VV)$ is compensated by a decrease in all the other channels shown.

\section{Experimental Constraints and Future Prospects}\label{s.Exp}

Having determined the regions of parameter space in which thermal freeze-out can account for the observed relic density, we now turn to other searches for dark matter.
We will discuss briefly the relevant characteristics of the different experiments before analyzing the impact of these observational efforts on the parameter space of thermal dark matter.

\subsection{LHC}\label{s.LHC}

Dark matter that primarily couples to top quarks suffers from a severely suppressed direct pair production cross section.  Therefore, the most promising way to probe this type of model is via the production of  the mediator $\tilde{t}$. Scalar top partners are charged under QCD, so pairs of the mediator can be produced copiously at the LHC, with the rate  depending only on the mass $m_{\tilde{t}} $. 
The signature of $\tilde{t}$ depends on the available decay channels which are primarily governed by the mass difference between the dark matter and the $\tilde{t}$. As long as $m_{\tilde{t}}-m_{\chi} \geq m_t$, the decay $(\tilde{t} \rightarrow \chi t)$ will be dominant, whereas in the range $ m_t \geq m_{\tilde{t}}-m_{\chi} \geq m_W + m_b$, the three-body decay, $(\tilde{t} \rightarrow \chi W b)$, becomes relevant. Therefore, the LHC limits on our model are similar to those on models for direct stop production with branching ratios of $100\%$ into these channels  and the limits derived by the ATLAS~\cite{Aad:2014nra, Aad:2014bva,Aad:2014qaa} and CMS~\cite{CMS:2014wsa,Chatrchyan:2014lfa,Chatrchyan:2013xna} collaborations apply without any modifications.  However, when  
$ m_W + m_b \geq m_{\tilde{t}}-m_{\chi}$, the region in which we are predominantly interested, the flavor-violating process, $(\tilde{t} \rightarrow \chi c)$, as well as the four body decay, $(\tilde{t} \rightarrow \chi b f \bar{f}  )$, could in principle contribute significantly.  In our model, absent additional flavor violation, the four body channel would dominate.  However, various UV completions to our simplified model could allow the relative contribution of these channels to be a free parameter without changing the cosmology.  Within the MSSM, the branching ratio into both the decay modes can be significant without violating flavor constraints, see e.g., Refs.~\cite{Blanke:2013uia,Agrawal:2013kha}.

Both ATLAS and CMS have analyses optimized for each of these channels, under the assumption of  100\% branching ratio into either decay mode~\cite {Aad:2014kra,CMS:2014yma}. At small mass splittings the monojet search is interpreted for both the channels, and found to have similar sensitivity.  This is unsurprising; at very small mass splittings, one would expect the final state particles to go undetected,  for either of the two decay channels.   
 As the mass splitting increases, particles in the four body decay might run afoul of lepton or jet vetoes in the present analysis.   A dedicated search with charm-tagging is relevant for the 2 body decay, and has sensitivity primarily when the mass splitting is larger than about 20 GeV \cite{Aad:2014kra}.   Also at larger mass splitting (but still less than $m_{W}$) if the 4-body decay dominates, there is some sensitivity to final states using  soft leptons~\cite{Aad:2014kra}.  In this moderate mass splitting regime,  it has been emphasized that the exclusion limits are in fact very sensitive to this branching ratio and can be weakened considerably when both channels have competing branching ratios~\cite{Grober:2014aha}.  In our analysis, we  impose limits obtained  from ATLAS, similar results would be obtained if the CMS limits were used instead.  We will display both a monojet region (which we expect to be insensitive to the branching ratio), and a region that explicitly relies on charm tagging.

In addition, with the given mass splittings between the bino and the stop, even though LHC14 can only probe up to $\approx$ 500 GeV in stop masses, the entire mass range consistent with the observed relic density can be comfortably probed by a 100 TeV collider~\cite{Low:2014cba}. These projections are for monojet searches, without any assumption about the decay mode, and so should be quite robust.\footnote{We reemphasize, that when the mass splitting approaches $m_W$, the sensitivity of the monojet search for the four-body decay interpretation will be reduced if the lepton and jet vetoes employed by the ATLAS and CMS collaborations for the current monojet searches are used.}

There can also be a significant impact on gluon fusion for the observed 125 GeV Higgs boson due to the presence of $\tilde{t}$. The gluon fusion amplitude can be simply written down using the low energy theorem~\cite{Kniehl:1995tn, Carena:2012xa, Carena:2013iba}: 
\beq\label{e:gf}
\mathcal{A}_{hgg} \simeq \mathcal{A}_{hgg}^{\rm SM} +  \frac{\lambda_h v^2}{2 m_{\tilde{t}}^2},
\eeq
where we have used the normalization that the SM contribution due to the top loop is $ \mathcal{A}_{hgg}^{\rm SM} $=4.\footnote{ A similar contribution is induced in the diphoton decay width of the Higgs due to the presence of $\tilde{t}$. However, the SM contribution due to the $W$ and $t$ loops has opposite sign and is much larger in magnitude, $\mathcal{A}_{h\gamma\gamma}^{\rm SM}=-13$~\cite{Ellis:1975ap, Shifman:1979eb, Gunion:1989we,Kniehl:1995tn,Carena:2012xa,Carena:2013iba}. Additionally, since the rate into $\gamma\gamma$ is proportional to the dominant production mode of the Higgs, gluon fusion, the total impact of a scalar top partner on the diphoton decay rate is further diluted.}
The above is an excellent approximation in the limit that the relevant $\tilde{t}$ mass is sufficiently heavier than the Higgs boson mass. 
The best fit signal strength reported by CMS for the 125 GeV Higgs boson from gluon fusion is $\mu_{hgg}=0.85^{+0.19}_{-0.16}$~\cite{Khachatryan:1979247}. Since scalar top partners without mixing can only give an enhancement, it is reasonable to impose the requirement that the contribution of $\tilde{t}$ to gluon fusion does not exceed 
the SM value by more than $20\%$. This implies a constraint on $m_{\tilde{t}}$ as a function of $\lambda_h$, which can be rewritten as a constraint on $r$ as a function of $m_\chi$ and $\lambda_h$:
\beq\label{eq:rgf} 
r \gtrsim \frac{v}{2 m_\chi}\left[\frac{\lambda_h}{2\left( \sqrt{\mu_{hgg}}-1\right)}\right]^{1/2} \sim 1.14 \sqrt{\lambda_h}\;\frac{ v}{m_\chi},
\eeq
assuming $\mu_{hgg}$ is bounded to be less than 1.2.
Clearly, as experimental precision and theoretical control increase, the above constraint will become stronger and, as we will see, is already quite restrictive for the case of large $\lambda_h$.

\subsection{Indirect Detection}\label{s:ID}

The annihilations $(\chi\chi\rightarrow t\bar t)$ are expected to occur today in regions with an overdensity of dark matter particles, thus leading to potentially observable signals in indirect detection experiments. In these regions, dark matter particles typically have very low velocities, therefore, the self-annihilation cross section is almost entirely $s$-wave, in contrast to the time of freeze-out, where both the $s$- and $p$-wave contribution were relevant. Using Eq.~(\ref{eq:GenOmegaNum}), in the absence of coannihilation, the annihilation cross section today for a thermally produced dark matter particle can be approximated by
\begin{equation}
\sigma {\rm v} \simeq \left(\frac{a}{a+3 b/x_{F}}\right) \left(\frac{x_F}{25}\right)
\left(\frac{g_{\ast}}{80}\right)^{-1/2}
\,3\times 10^{-26}\mbox{cm}^3/\mbox{s}\;,
\end{equation}
where $a$ and $b$ are given in Eqs.~(\ref{eq.chitt-a}) and (\ref{eq.chitt-b}), respectively, and $x_{F}\sim 25$. In particular, we find that $\sigma {\rm v}$ ranges between $(2-0.1)\times 10^{-26} \, \mbox{cm}^3/\mbox{s}$ for $m_\chi =250~ \GeV-  2 \TeV$. These values of the cross section lie below the present upper limits from H.E.S.S.~\cite{Abramowski:2011hc} and Fermi-LAT~\cite{Ackermann:2013yva}, however, depending on the dark matter density profile in the galactic center, the prospected Cherenkov Telescope Array (CTA), might possess the sensitivity necessary to probe $\sigma {\rm v} \geq 1\times 10^{-26}\mbox{cm}^3/\mbox{s}$  in this mass region~\cite{Silverwood:2014yza}.  In the coannihilation region, on the other hand, the cross section for $(\chi\chi\rightarrow t\bar t)$ is highly suppressed, therefore the detection of annihilation signals will be very challenging in this regime.

Recently, after a reevaluation of the background uncertainties and a new derivation of the gamma ray spectrum \cite{Calore:2014xka}, 
$(\chi\chi \to t\bar{t})$  has also been considered as a possible explanation of the Galactic Center gamma-ray excess \cite{Agrawal:2014oha,Calore:2014nla}.  Interestingly, it was found that for $m_\chi \lesssim 200~\GeV$, annihilation of  thermal bino dark matter into top pairs might account for the observed excess, even though the $p$-value is rather low \cite{Calore:2014nla}.

\subsection{Direct Detection}

At first sight the prospects for the observation of dark matter coupling to top quarks do not seem to be particularly encouraging as the absence of top quarks in the nucleus prevents tree-level interactions. However, we find that loop diagrams can change this picture considerably and can
induce a spin-independent dark matter nucleus scattering cross section within the reach of upcoming experiments.

\begin{figure}
\begin{tabular}{ccc}
 \includegraphics[width=0.25\textwidth]{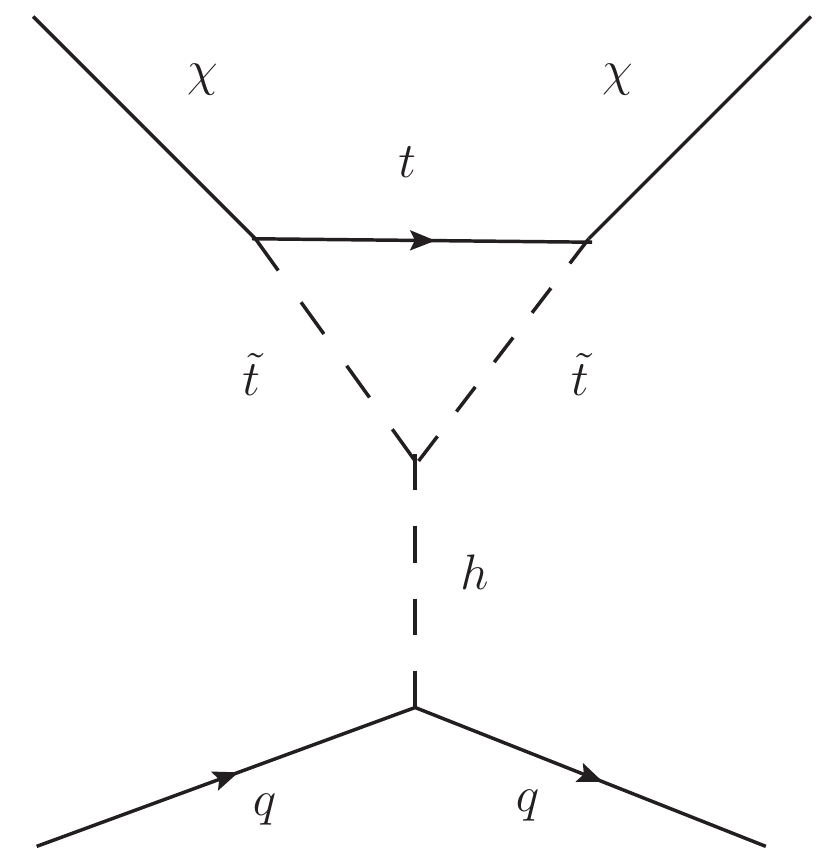}~~ & ~~\includegraphics[width=0.25\textwidth]{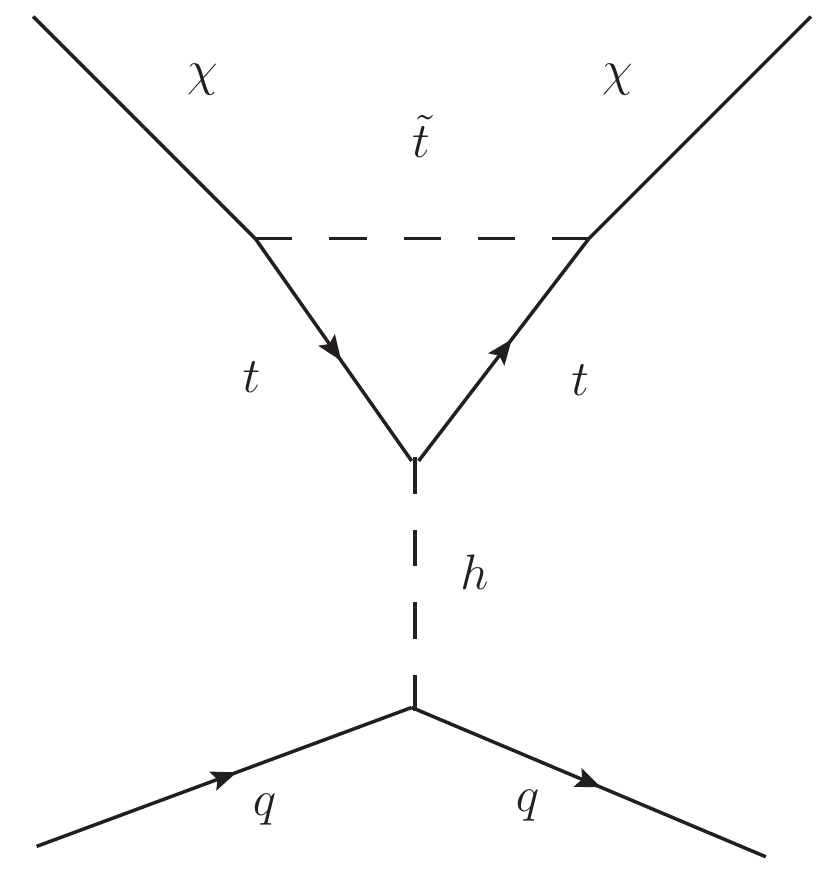}~~ & ~~
 \includegraphics[width=0.27\textwidth]{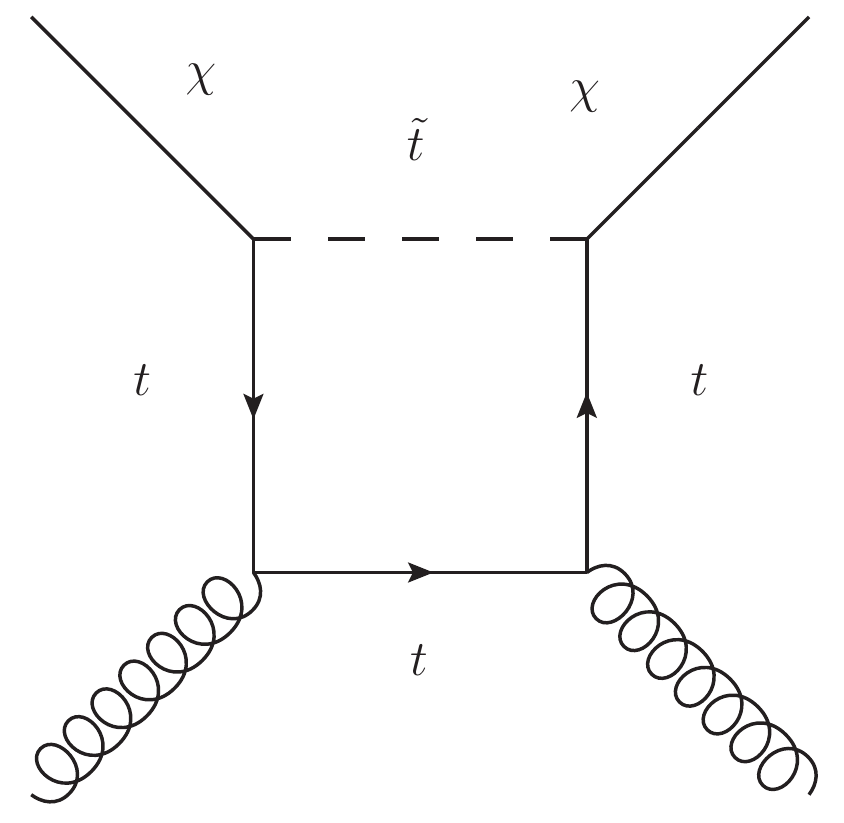} \\
 (a) & (b) & ~~~(c)
\end{tabular}
\caption{ \label{fig:DDdiagrams} { \em Representative examples of the triangle ((a) and (b)) and  (c) box diagrams which contribute to the dark matter nucleon coupling.}}
\end{figure}

The first process which can generate a sizable direct detection cross section is due to a loop induced coupling of the dark matter with the Higgs boson. The effective $\chi\chi h$ coupling is generated by triangle diagrams with scalar top partners and top quarks in the  loop, see Figs.~\ref{fig:DDdiagrams}a and \ref{fig:DDdiagrams}b for diagrams. This effective coupling has been calculated in the case of supersymmetry \cite{Djouadi:2001kba} and the results can be applied to our model with the appropriate replacements. We re-derived the dark matter Higgs boson coupling using the low energy Higgs theorem \cite{Kniehl:1995tn} and find good agreement between our calculation and the result of Ref.~\cite{Djouadi:2001kba}. Loop induced interactions with the $Z$, which are known to be important for Dirac dark matter interacting with tops \cite{Batell:2013zwa}, are not expected to be relevant here as vector interactions vanish for Majorana fermions.

The second relevant loop effect is due to the fact that the dark matter can couple to the gluon content in the nucleus via a box diagram with scalar mediators and top quarks in the loop~(e.g. Fig.~\ref{fig:DDdiagrams}c)~\cite{Drees:1993bu}. 

Another effect that cannot be neglected when $\lambda_h$ is large, is due to the loop induced coupling of the Higgs boson to gluons via $\tilde{t}$. However, this will be a two-loop effect in the direct dark matter detection cross-section. Furthermore, this is precisely the loop which leads to modifications of the $gg \rightarrow h$ rate, and hence the region where this would have a relevant effect on the direct dark matter detection cross-section would lead to an unacceptably large deviation in the production of the Higgs boson at the LHC. Therefore, we do not consider this effect in our analytics below. However, it is always included in our full numerical calculations which were computed using the built-in analytical MSSM formulae in {\tt MicrOmegas 3.3}, with appropriately rescaled couplings, including the effective $\chi\chi h$ coupling computed according to Ref.~\cite{Djouadi:2001kba}.

The direct detection cross-section per nucleon is given by:
\be
\sigma_{SI}^n = \frac{4m_\chi^2m_n^4\, \mathcal{A}^2}{\pi (m_\chi+m_n)^2}\, ,
\ee
where $n$ refers to either the neutron or the proton and $\mathcal{A}$ is the amplitude. $\mathcal{A}$ can be written as follows:
\be
\mathcal{A}=\mathcal{A}^h+\mathcal{A}^g,  
\ee
and we will further decompose the contributions in the amplitude due to the exchange of the Higgs boson in a part which depends on $\lambda_h$ (where the $\tilde{t}$ couples to the Higgs boson, Fig.~\ref{fig:DDdiagrams}a)  and independent of $\lambda_h$ (where the $t$ couples to the Higgs boson, Fig.~\ref{fig:DDdiagrams}b):
\be
\label{eqn:HiggsExchangeDD}
\mathcal{A}^h= \frac{\left( \mathcal{A}^h_t+\mathcal{A}^h_{\tilde{t}} \right)}{2 v m_h^2} \sum_q f_q^n, \qquad\qquad f_q^n = \left \{\begin{array}{c c}  f_{Tq}^n,& \quad  q=\{u,d,s\} \\
  \frac{2}{27} f_{Tg}^n, & \quad q=\{c,b,t\} 
    \end{array} \right. \,, 
    \ee
where $f_{Tg}^n = (1-\sum_{u,d,s}f_{Tq}^n)$ and  $f_{Tq}^p = \{  0.0153, 0.0191, 0.0447\}$, $f_{Tq}^n = \{0.0110, 0.0273, 0.0447\}$ are the default values used in {\tt MicrOmegas 3.3}, leading to $\sum_q f_q^n \simeq 0.28$. 
    
 As the full analytic expressions for both the effects described above are lengthy and cumbersome we do not repeat them here and refer the reader to Refs.~\cite{Djouadi:2001kba, Drees:1993bu}. However, certain expansions can be made in the parameter regions of interest which gain us insight in the behavior of the various contributions. 
 
From our numerical results in Sec.~\ref{Relic.Numb}, we know that the mass splitting required in the MSSM to obtain the experimentally consistent relic density is less than $\sim 45$ GeV, for $m_\chi \lesssim 2 $ TeV. When we allow $y_\chi$ and $\lambda_h$ to be free parameters, the mass splitting can be significantly larger. In fact, for very heavy dark matter masses with appreciable mass splitting, 
we will see that the direct detection cross-section can be enhanced, partially due to the fact that a large $y_\chi$ is needed to obtain a consistent relic density. Therefore, we will present analytical expressions for two regimes:  $\Delta m$ smaller or larger than the top mass. 

When the mass splitting between $m_\chi$ and $m_{\tilde{t}}$ is smaller than $m_t$, we can expand in the small parameters $\delta=(m_t-m_{\tilde{t}}+m_\chi  )/m_\chi$ and $\Sigma= (m_t+m_{\tilde{t}}-m_\chi)/m_\chi$:   
\bea
  \mathcal{A}_t^h&\simeq&\frac{3  y_{\chi }^2 }{4 \pi ^2 }  \frac{m_t^2}{v
   m_{\chi }}\left[1-\frac{1}{4} (1-\delta ) (\Sigma +1) \log
   \left[\frac{m_{\tilde{t}}^2}{m_t^2}\right]-\frac{\delta}{3 \Sigma }  (2+5 \Sigma )+\frac{\delta ^2 }{15
   \Sigma ^2}(4+11 \Sigma)\right]\, , \nonumber\\
   && \label{eq.DDtopsmall}\\
  \mathcal{A}_{\tilde{t}}^h &\simeq& \frac{3  \lambda _h y_{\chi }^2}{16 \pi ^2 } \frac{v}{m_{\chi }} \left\{-1+\frac{\Sigma}{2}   \left(1-\frac{\delta
   }{\Sigma }-\delta \right) \log
   \left[\frac{m_{\tilde{t}}^2}{m_t^2}\right] \right.  \nonumber \\
   && \qquad \qquad \qquad  \left. -\frac{\Sigma}{2}  
   \left[1-\frac{\Sigma }{2}-\frac{\delta}{3 \Sigma
   }  \left(19 +\frac{2 \Sigma }{3}\right)+\frac{\delta ^2}{5 \Sigma
   ^2} \left(16+\frac{27 \Sigma }{2}\right)\right]\right\}  \, , \nonumber \\
   &&\label{eq.DDstopsmall}\\
  \mathcal{A}^g&\simeq& \frac{ y_{\chi }^2 }{4  m_{\chi }^3 \, \Sigma ^2}\left[\frac{f_{Tg}}{135}+\frac{G_{\alpha }^t }{8
   \pi }\left(\frac{3 \Sigma ^2}{4 \pi } \log
   \left[\frac{m_{\tilde{t}}^2}{m_t^2}\right]+\frac{13}{15}\right)\right] \, , \label{eq.DDglusmall}
\eea
where the default value for $G_{\alpha}^t= $ 0.053 from {\tt MicrOmegas 3.3}.  Note that when the mass splitting, $m_{\tilde{t}}-m_\chi$, is much much smaller than $m_t$, then $\delta \sim \Sigma$, however, when the mass splitting is close to the top mass, then $\delta << \Sigma$. We have checked the above expressions reproduce the full numerical results within 30\% for all three amplitudes when $m_\chi \gtrsim 500 $ GeV, leading to an estimation for the total cross-section which is accurate to 50\%. For smaller $m_\chi$, the above approximation gives results within a factor of 2 from the full numerical calculation as long as $r$ is not too small ($\delta$ small).

Typically one expects the Higgs boson exchange to dominate the dark matter nucleon coupling, however, as can be seen from Eqs.~(\ref{eq.DDtopsmall})-(\ref{eq.DDglusmall}), all the different amplitudes scale with approximately $1/m_\chi$ in this regime. In particular, this means that the dark matter coupling to gluons can become comparable to its coupling to the Higgs boson in this small mass splitting regime, especially for dark matter mass, $m_{\chi}$, 
close to $m_t$. Unfortunately, the contributions from the triangle and the box diagrams interfere destructively such that direct detection experiments have reduced  sensitivity to dark matter with $m_\chi$ close to $m_t$.

When the mass splitting is larger than the top mass, one can take the limit that the mass of the scalar top partner is much larger than both $m_\chi$ and $m_t$. In this case, the various contributions to the amplitude are substantially simplified: 
    \bea
  \mathcal{A}_t^h&\simeq&  -\frac{3\,y_\chi^2}{16\pi^2}\frac{m_\chi m_t^2}{v m_{\tilde{t}}^2}\left\{ 1+\frac{2m_\chi^2}{m_{\tilde{t}}^2} \left[\frac{1}{3}+ \frac{m_t^2}{ m_\chi^2}\left(\frac{3}{2}-\log\left[ \frac{m_{\tilde{t}}^2}{m_t^2}\right]   \right)\right]\right\} \, , \label{eq.DDtopbig}\\
  \mathcal{A}_{\tilde{t}}^h &\simeq& -\frac{ 3\, \lambda_h\, y_\chi^2}{32 \pi^2}\frac{v\, m_\chi}{m_{\tilde{t}}^2}\left[ 1+\frac{m_\chi^2}{ m_{\tilde{t}}^2}\left( \frac{1}{3}-\frac{m_t^2}{m_\chi^2} \right)  \right]\, , \label{eq.DDstopbig}\\
  \mathcal{A}^g&\simeq& y_\chi^2 \frac{m_\chi}{m_{\tilde{t}}^4}\left\{ \frac{f_{Tg}}{108}+\left(\frac{1}{32}\log\left[\frac{m_{\tilde{t}}^2}{m_t^2}\right]-\frac{9}{128}\right)\frac{G_\alpha^t}{\pi} \right\} \, .  \label{eq.DDglubig}
\eea
Using the above expansions, 
the Higgs exchange amplitude $\mathcal{A}^h$, defined in Eq.~(\ref{eqn:HiggsExchangeDD}), goes as 
 $ 1 /(m_h m_{\tilde{t}})^2$ whereas
  $\mathcal{A}^g$ is proportional to $1/m_{\tilde{t}}^4$. Consequently, the contribution from the triangle diagrams will always dominate the cross section for large $m_{\tilde{t}}$. 
 Comparing the above amplitudes with the full numerical results, the contribution from $\tilde{t}$, $\mathcal{A}_{\tilde{t}}^h$, is within 10\% of the full numerical amplitude, even for small $r\sim 1.2$ and $m_\chi \gtrsim 500$ GeV. The top contribution, $\mathcal{A}_t^h$, is accurate to about 20\% across the region of interest. The contribution from the box diagrams with the gluon $\mathcal{A}^g$ is only accurate to approximately an order of magnitude,
  but the gluon contribution is negligible in this regime. 
  Even ignoring it completely, the direct detection cross-section is within a factor of 2 of the full numerical calculation.

\subsection{Numerical Results}

In the following we discuss the impact of various experimental probes 
described above on the allowed parameter space of thermal dark matter.  We begin with the description of the effect on the more general simplified model.  We also draw conclusions  for
our minimal MSSM scenario.

As discussed in the previous subsection, in the simplified effective stop model, the loop induced couplings between the dark matter and the SM
can have a significant impact on the direct detection cross-section. In addition, as pointed out in Sec.~\ref{s:ID}, in the blue shaded region shown in Figs.~\ref{fig:InStatesl1} and \ref{fig:InStatesl4}, annihilations could give rise to detectable signals at future indirect detection experiments~\cite{Silverwood:2014yza}. The measurement of the  
gluon fusion cross section at the LHC is sensitive to the value 
$\lambda_h$ and excludes a large portion of the parameter space under consideration for the largest value we consider. We stress again that direct searches at the LHC do not depend on either of the couplings, but are only sensitive to the mass splitting between the two states and the overall mass scale.

\begin{figure}[ptb]
{\includegraphics[width=4.25in, angle=0]{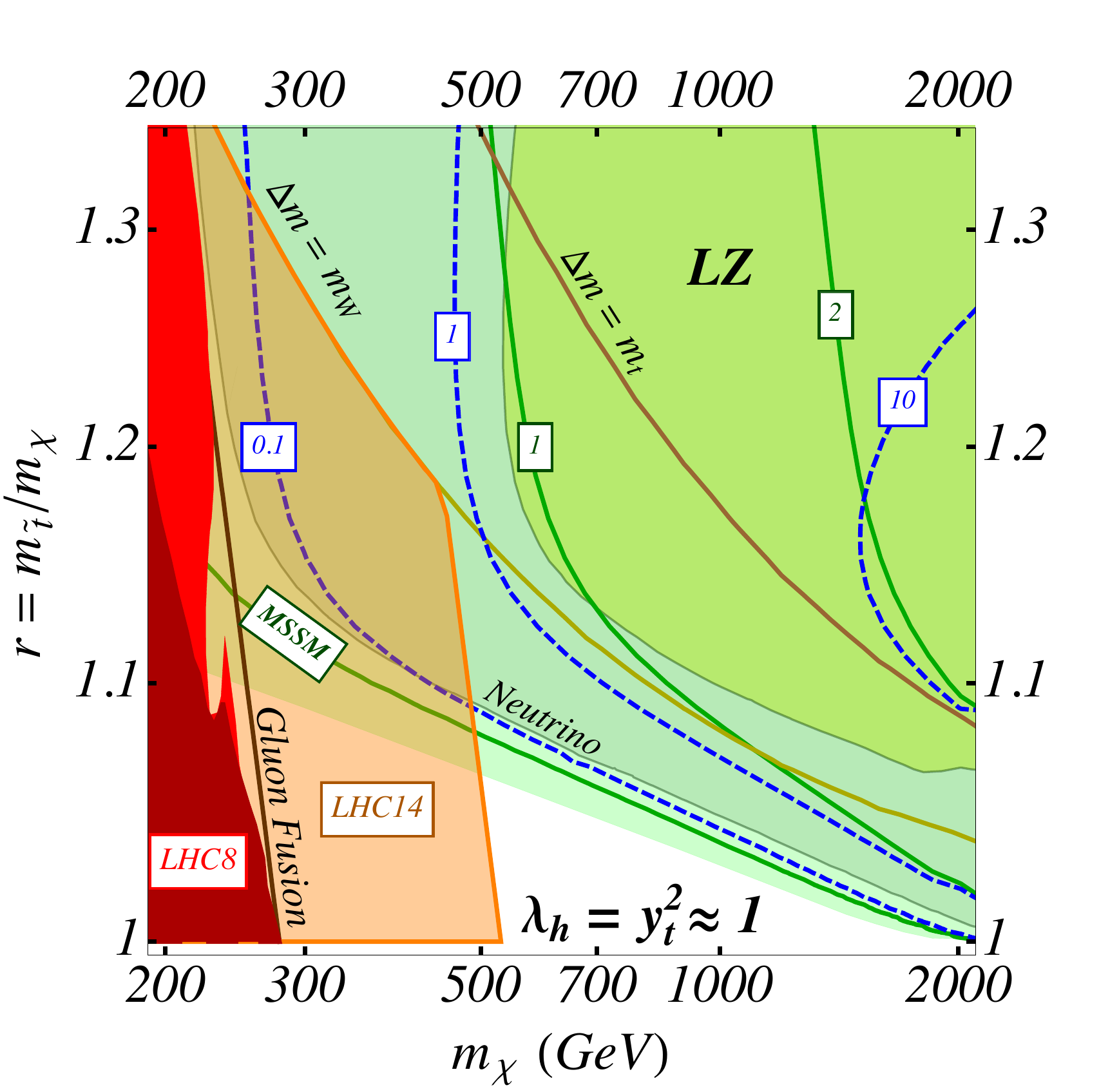}}
\caption{\label{fig:Everything_lambda1}{\em 
The blue dashed contours label direct detection cross-section in units of $10^{-47} \;\rm cm^2$. Green contour lines show the value of $y_\chi$ required to saturate the relic density. The green contour labeled ``MSSM" corresponds to $y_{\chi} = \frac{2}{3} \sqrt{2}g_{Y}$, and lies below the irreducible neutrino background, denoted as ``Neutrino"~\cite{Cushman:2013zza}.  The bright green shaded region corresponds to the region which will be probed by  LZ~\cite{Cushman:2013zza}. The region not shaded green is where it is not possible to saturate the relic density constraint thermally in  this scenario.  Also displayed are lines showing where the mass splitting $\Delta m = m_{\tilde{t}}-m_\chi = m_W \; {\rm or} \; m_t $. Red~(dark: monojet, bright: charm-tagging) and orange~(monojet) regions denote current~\cite{Aad:2014nra,Aad:2014kra} and projected~\cite{Low:2014cba} exclusion bounds from the LHC.  The region to the left of the black line denoted as ``Gluon Fusion" would give rise to more than 20\% enhancement in gluon fusion compared to the SM expectation, Eq.~(\ref{eq:rgf}). 
 }}
\end{figure}

\begin{figure}[tbp]
\subfloat[]{\includegraphics[width=4.1in, angle=0]{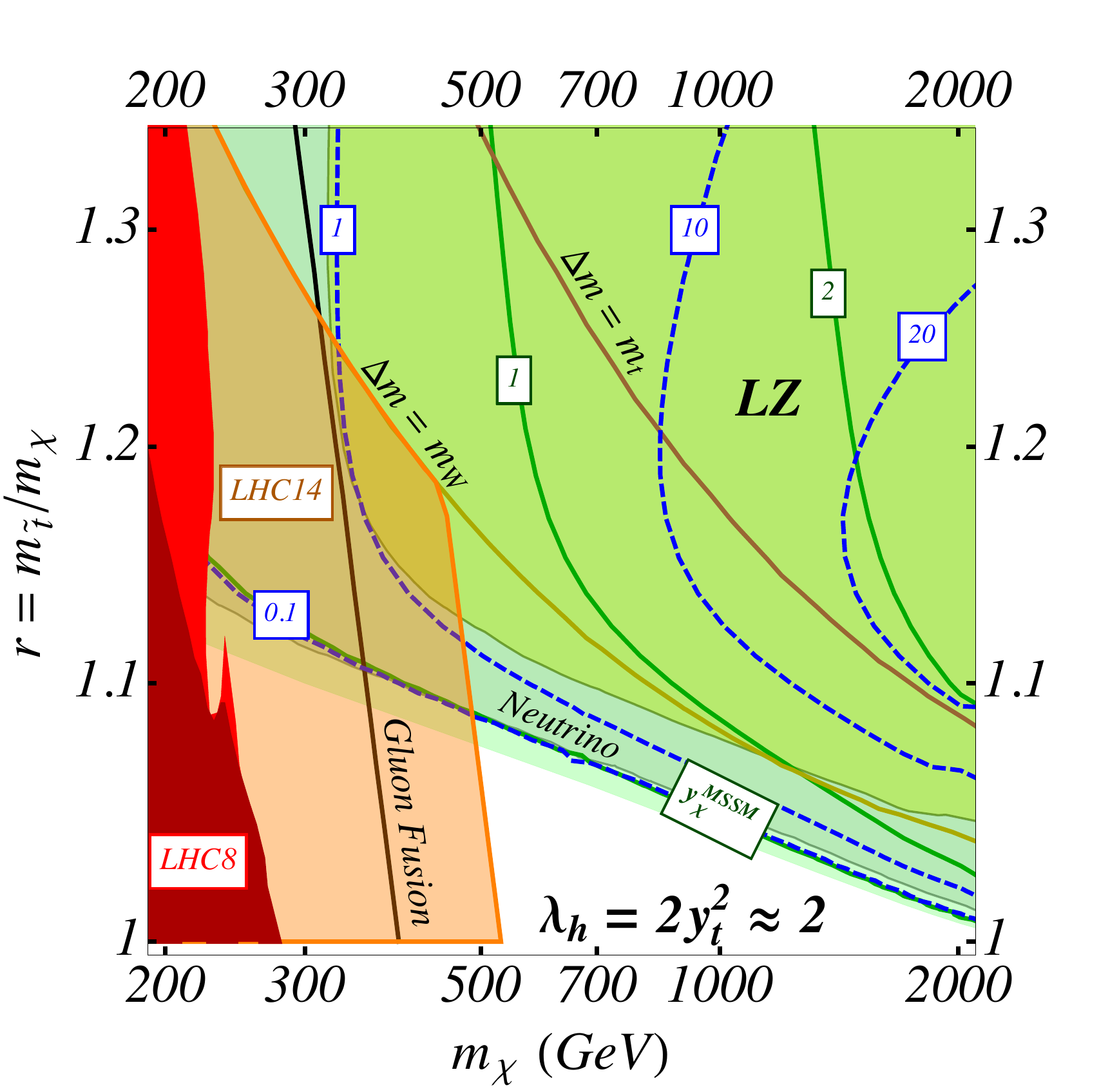}\label{fig:Everything_lambda2}}~~\\
\subfloat[]{\includegraphics[width=4.1in, angle=0]{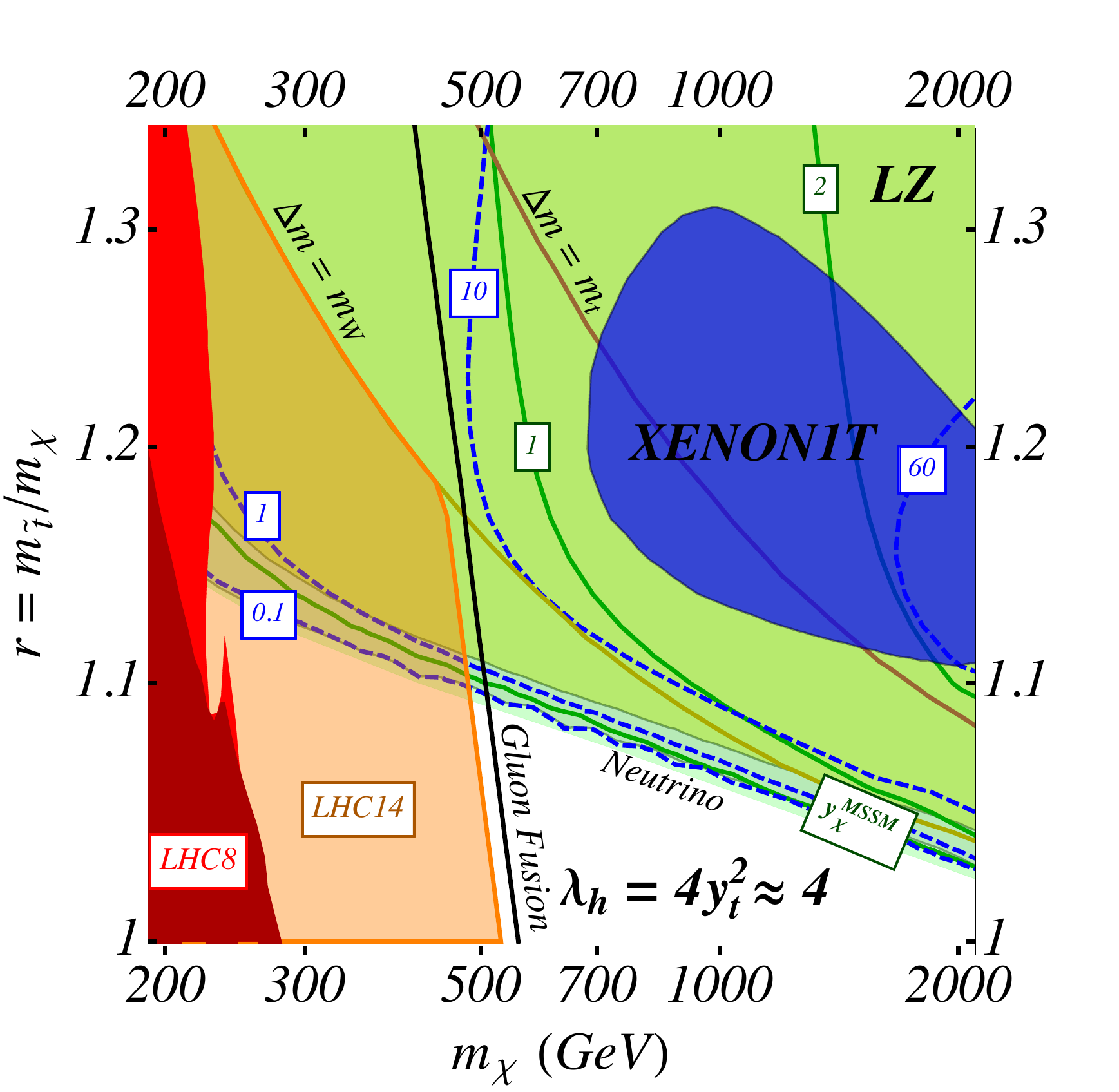}\label{fig:Everything_lambda4}}
\caption{\label{fig:Everything_lambda24}{\em Same as Fig.~\ref{fig:Everything_lambda1}, but showing results for (a) $\lambda_h=2y_t^2\approx 2$ and (b) $\lambda_h=4y_t^2\approx 4$. The blue shaded region in (b) corresponds to the region that will be probed by XENON1T~\cite{Cushman:2013zza}.  }}
\end{figure}

Figs.~\ref{fig:Everything_lambda1} and \ref{fig:Everything_lambda24} summarize our results for the $\lambda_h\approx1$, 2 and 4 cases respectively. Green contour lines show the value of $y_\chi$ required to saturate the relic density. The green contour labeled ``MSSM" in Fig.~\ref{fig:Everything_lambda1} corresponds to the minimal supersymmetric model for our set up, discussed in Sec.~\ref{s:SUSY}, given by $y_{\chi} = \frac{2}{3} \sqrt{2}g_{Y}$ (also denoted in Fig.~\ref{fig:Everything_lambda24} as the contour labeled $y_\chi^{MSSM}$). The region not shaded green is where it is not possible to saturate the relic density constraint thermally in  this scenario.  The blue dashed contours label direct detection cross-section in units of $10^{-47} \;\rm cm^2$. As a comparison we indicate the so called neutrino floor, labeled as ``Neutrino", where the rate of dark matter interactions in a detector equals the rate of recoil events induced by coherent scattering of neutrinos  (primarily atmospheric neutrinos for the masses we consider). As these neutrino interactions constitute an irreducible background, direct detection experiments cannot be expected to probe appreciably smaller cross sections. This irreducible background interpolates between $\sim 4\times10^{-49} - 4 \times 10^{-48} $~cm$^2$ for the mass range considered~\cite{Cushman:2013zza}. Except for the $\lambda_h\approx 4$ case, the direct detection cross-section for $y_\chi^{MSSM}$ lies mostly below the neutrino background floor. The bright green shaded region corresponds to the region which will be probed by  LZ~\cite{Cushman:2013zza}.  XENON1T will  have sensitivity in the $\lambda_h\approx 4$ case, which is depicted by the shaded blue region~\cite{Cushman:2013zza}. 

Also shown are lines indicating where the mass splitting $\Delta m = m_{\tilde{t}}-m_\chi = m_W \; {\rm or} \; m_t $. Darker red and red regions denote current LHC exclusion regions due to monojet and charm-tagging respectively~\cite{Aad:2014nra,Aad:2014kra}. The orange region denotes the projected LHC14  monojet exclusions~\cite{Low:2014cba}. Recall that the exclusions due to the monojet searches are expected to be model-independent; however, the exclusion limits obtained via charm-tagging  could be significantly weakened as discussed in Sec.~\ref{s.LHC}. The region to the left of the black line denoted as ``Gluon Fusion" would give rise to more than 20\% enhancement in gluon fusion compared to the SM expectation, Eq.~(\ref{eq:rgf}). Outside of the region plotted in Figs.~\ref{fig:Everything_lambda1} and \ref{fig:Everything_lambda24} are additional LHC constraints for $\Delta m>m_t$ in the low $m_\chi$, large $r$ region. Specifically, for $\Delta m\sim m_t$, even though current bounds only extend to $m_\chi \sim 275$ GeV~\cite{Aad:2014bva,Aad:2014kra, Aad:2014qaa}, LHC14 is expected to probe $m_\chi \sim 500$ GeV~\cite{CMS:2013xfa}. These searches will not probe the coannihilation region, but will be complemented by possible signatures in future indirect detection experiments~\cite{Silverwood:2014yza}.  

We now discuss the direct detection contours in some detail. We recall that the direct detection cross-section is always proportional to $y_\chi^4$, which we fix by requiring a consistent relic density at every point. This requirement implies that for every $m_\chi$ there is a maximum occurring at some $r$ and $\sigma_{SI}$ is increasing with $m_\chi$. To see this we first note that coannihilations begin to suppress the required $y_\chi$ severely for $r\lesssim 1.1$ (c.f. Fig.~\ref{fig:gBRR}), and so the smallest direct detection cross-sections are found in the  most degenerate region.  The maximum occurs approximately when the coannihilation processes become irrelevant for setting the relic density, and the relevant process is $(\chi\chi \to t\bar{t})$~(corresponding to when the contribution of this channel is $\gtrsim 70\%$). This occurs for $r \sim 1.15-1.3$ across the mass range we consider. For larger $r$, $y_\chi^4$ can be determined by examining the partial wave expansion for the $(\chi \chi \rightarrow t \bar{t})$ process. In particular, $y_\chi^4$ is inversely proportional to $(a+3b/x_F)$, with values set as in  Eqs.~(\ref{eq.chitt-a}) and (\ref{eq.chitt-b}):  
\beq\label{eq:yscale}
y_\chi^4 \propto \left[\frac{3 m_t^2}{32   \pi  \left(r^2+1-m_t^2/m_\chi^2\right)^2 m_{\chi }^4}+\frac{3 \left(r^4+1\right)}{16 \pi    \left(r^2+1\right)^4  m_{\chi }^2\, x_F}\right]^{-1} .
\eeq
One expects that in the region where the $s$-wave contribution is dominant, $y_\chi^4$ scales as $m_\chi^4$ and as $m_\chi^2$ in the region where the $p$-wave contribution is the most relevant. Due to the different $r$ dependence of the $s$ and $p$ wave contributions, the annihilation cross-section is not dominantly $p$-wave until almost $m_\chi\sim$ 2 TeV.  

Now turning to the scaling of $\sigma_{SI}$ with $m_\chi$ and $r$, we note that in the small mass splitting regime, Eqs.~(\ref{eq.DDtopsmall})-(\ref{eq.DDglusmall}),
\beq\label{eq:sigmascalesmall}
\sigma_{SI}^{\Delta_m <m_t} \propto \frac{y_\chi^4}{m_\chi^2}\, .
\eeq
Instead in the large mass splitting regime, Eqs.~(\ref{eq.DDtopbig})-(\ref{eq.DDglubig}),
\beq\label{eq:sigmascalebig}
\sigma_{SI}^{\Delta_m >m_t} \propto \frac{y_\chi^4}{r^4 m_\chi^2}\left(1+\frac{1}{r^2}\right)^2 \, .
\eeq
Comparing Eqs.~(\ref{eq:yscale})-(\ref{eq:sigmascalebig}), we see clearly that for a fixed $r$, if the cross-section setting the relic density is predominantly $s$-wave, then the direct detection cross-section increases with $m_\chi$: $\sigma_{SI}\propto m_\chi^2$. When the relic density is instead set by the $p$-wave contribution, the dominant scaling of the annihilation cross-section and the direct detection cross-section are the same, and we expect very little sensitivity to increasing either $m_\chi$ or $y_\chi$. Consequently, the direct detection cross section of thermal dark matter increases with $m_\chi$ in the mass range of interest. 
 
The stronger variation with $r$ of $\sigma_{SI}$ for increasing $m_\chi$ can also be understood by comparing the $r$ dependance in Eq.~(\ref{eq:sigmascalesmall}), using the first term for $y_\chi^4$ in Eq.~(\ref{eq:yscale}) (small mass splitting and $s$-wave dominated annihilation cross-section) with the $r$ dependance obtained in Eq.~(\ref{eq:sigmascalebig}) using the second term for $y_\chi^4$ in Eq.~(\ref{eq:yscale}) (large mass splitting and $p$-wave dominated annihilation cross-section). 

We also comment briefly on the scaling with $\lambda_h$: Since the leading dependance of both $\mathcal{A}_t^h$ and $\mathcal{A}_{\tilde{t}}^h$ with $m_\chi$ and $r$ is approximately the same and $\mathcal{A}_g$ is  only relevant in small regions of parameter space, the direct detection cross-section approximately behaves as $(y_t^2+\lambda_h)^2$.

It is interesting to note the complementarity in the reach of the LHC and direct detection experiments. As mentioned previously, the direct search bounds from the LHC do not depend on the exact value of the couplings, but only on the mass splitting and the mass scale. The LHC14 should be able to probe masses up to 500 GeV in the $\Delta m <m_W$ region, a 100 TeV collider would be able to comfortably probe masses up to $\sim 2$ TeV~\cite{Low:2014cba}. On the other hand, both the enhancement in gluon fusion and the direct detection cross-section are impacted by  the mass splitting and the couplings.  For $\lambda_h\approx 1$, the current direct search limits from LHC and bounds from gluon fusion are approximately comparable. The direct detection cross-section is very suppressed in the region where the LHC bounds are expected to be strongest.  On the other hand, for the larger $m_\chi$, $r$, region, where LHC searches will have no sensitivity,  the large $y_\chi$ required to obtain an experimentally consistent relic density enhances the direct detection rate, which is  largest in this region. As can be seen in Fig.~\ref{fig:Everything_lambda1}, LZ is expected to probe $m_\chi\gtrsim 600$ GeV and $\Delta m \gtrsim m_W$. Once $\lambda_h$ is increased to 2, the constraint due to the enhancement of gluon fusion becomes stronger, however, the 14 TeV LHC is still expected to have stronger sensitivity. The direct detection cross-section also increases, and we can see from Fig.~\ref{fig:Everything_lambda2} that now there is an overlapping region between $m_\chi = 300$ and 500 GeV where both LZ and LHC14 will be sensitive. When $\lambda_h$ is pushed to an even larger value of 4,  the requirement of not having more than a 20\% enhancement in gluon fusion constrains a large region of parameter space for $m_\chi \lesssim 500$ GeV, somewhat stronger than the direct search sensitivity expected from LHC14. In addition, as can be seen in Fig.~\ref{fig:Everything_lambda4}, the direct detection cross-section is enhanced significantly such that certain regions will be accessible in the near future to XENON1T. 

\begin{figure}[t]
\includegraphics[width=4.1in, angle=0]{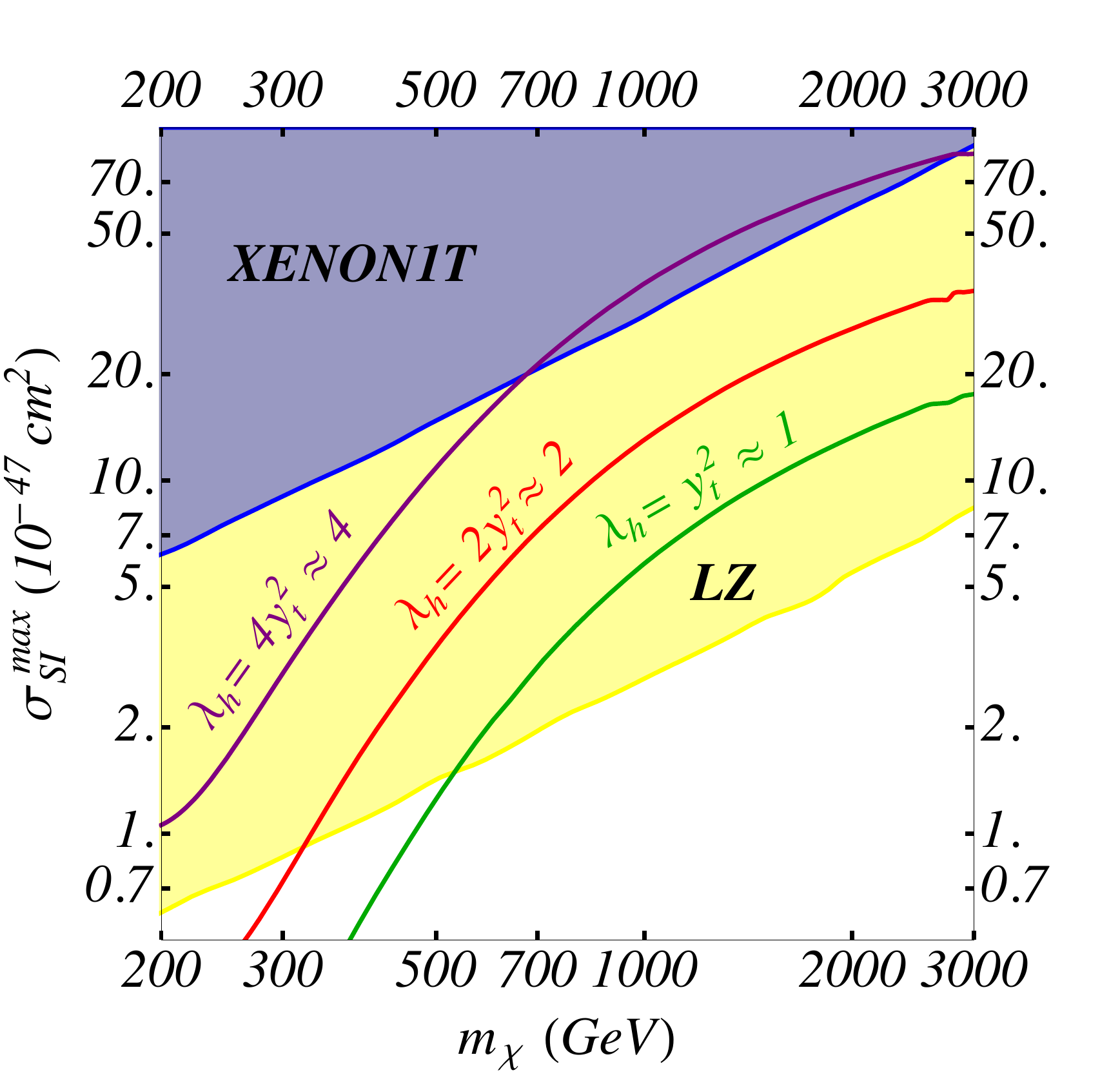}
\caption{\label{fig:DD_max}The maximum value of the dark matter-nucleon cross section for different values of $\lambda_{h}$ for points consistent with a thermal relic density.  Also shown are the expected sensitivities of both the XENON1T and LZ experiments.  We have imposed a maximum value of $y_{\chi} < 3$.  The exact value of this bound is largely irrelevant for determining the maximum cross section, but can effect the maximum cross section, particularly at small masses.  Moreover, it does determine the maximum $m_{\chi}$ for which a relic density may be successfully obtained.  For details, see the text.}
\end{figure}

Turning now to Fig.~\ref{fig:DD_max}, we have plotted the maximum achievable dark matter nucleon cross section in our model for three different values, $\lambda_h \approx 1$, $2$, and $4$, as a function of $m_{\chi}$.  For each value of $m_{\chi}$, we scan values of $m_{\tilde{t}}$ and $y_{\chi}$ consistent with the relic density and find the largest cross section. We reiterate that these maximal cross sections are typically found where the process $(\chi \chi \rightarrow t \bar{t})$ starts to dominate the cosmology (coannihilations are unimportant). Note that  $r$ is relatively constant for the maximal $\sigma_{SI}$ values, varying between $r\sim 1.15-1.3$, as can be seen from inspecting the direct detection contours in Figs.~\ref{fig:Everything_lambda1}-\ref{fig:Everything_lambda4}. For the lower dark matter masses, $m_\chi \lesssim $TeV, as stated earlier, the direct detection cross-section increases as $m_\chi^2$. However, as the annihilation cross-section transitions to be dominantly $p$-wave, the direct detection amplitude and early universe annihilation both approximately scale as  $\sim {y_{\chi}^{4}}/{m_\chi^2}$.  Thus, once the relic density is fixed, the direct detection cross section does not change dramatically even as $y_{\chi}$ and $m_\chi$ are increased. 

Nevertheless,  we do cut off the scan at values of $y_{\chi} < 3 \equiv y_{\chi}^{max}$ which corresponds to a dark matter mass of 3 TeV.  For most masses, the value of  $ y_{\chi}^{max}$ does not affect the maximum direct detection cross section -- it is determined by cosmology alone. However, at the lowest masses $(\lsim$ 250 GeV) it would be possible to raise the direct detection cross section very modestly (say 10\%) by allowing a larger $y_{\chi}^{\max} \approx 5$.     

An important take home message is that in WIMP models where the dominant interactions are with the top partners, cosmological considerations indicate a cross section well below current direct detection bounds.  Indeed, even XENON1T will have difficulty probing much of the parameter space.  LZ, however, will be much more effective.

While we have been considering this simplified model independent of the MSSM, it is worthwhile to ask what the above results imply for a MSSM spectrum that mimics the one we consider here.  In addition to the loop mediated contribution, important contributions to direct detection may arise from tree-level Higgs boson exchange which are generated by non-zero Higgsino-bino mixing.  Assuming that the non-Standard Higgs boson contribution decouples, the Higgs boson mediated direct detection of Eq.~(\ref{eqn:HiggsExchangeDD}) is modified to
\begin{eqnarray}
\label{eqn:HiggsExchangeDDtree}
\mathcal{A}^h= \frac{\left( \mathcal{A}^h_t+\mathcal{A}^h_{\tilde{t}} +  \mathcal{A}^{tree} \right)}{2 v m_h^2} \sum_q f_q^n \; ,
\end{eqnarray}
with
\begin{equation}
\mathcal{A}^{tree}  \approx - \frac{2 m_{Z}^{2} \sin^2{\theta_{W}} ( m_{\chi}  + \mu \sin{ 2 \beta} )}{v ( \mu^2 - m_\chi^2)}\,,
\end{equation}
where $\mu$ is the supersymmeteric Higgs mass parameter,  $\tan\beta=v_u/v_d$ is the ratio of the two MSSM Higgs field vacuum expectation values and $v_u^2+v_d^2=v^2$.

In the limit where only $\mathcal{A}^{tree}$ contributes, the relevant cross section may be approximated as:
\be
\sigma_{SI} \approx 3\times 10^{-47}\textrm{cm}^2 \left( \frac{1 \textrm{ TeV}}{\mu} \right)^4 \left( \frac{m_\chi}{500 \textrm{ GeV}} \right)^2 \left(1+\frac{\mu \, s_{2 \beta }}{m_{\chi }}\right)^2 \left(1-\frac{m_{\chi }^2}{\mu ^2}\right)^{-2}.
\ee
Looking back to Fig.~\ref{fig:Everything_lambda1},  it is easily possible  -- even for relatively large $\mu$ -- that this contribution can dominate the loop mediated one, which for the MSSM case lies at the 10$^{-48}$ cm$^2$ level (Note the cosmology would be relatively insensitive to the presence of a small Higgsino admixture.).   So in the case of a bino-$\tilde{t}_R$ MSSM-like case ($y_\chi \approx 2 \sqrt{2} g_Y/3$, $\lambda_{h} \approx 1$) an observed signal at LZ could be a hint of new dynamics (e.g. Higgsinos) at near the TeV scale.  Note a more general MSSM with stop mixing could effectively allow larger stop-Higgs couplings (due to large $A$-terms), which can modify the direct detection and cosmology in important ways.  This is the subject of upcoming work.

\section{Conclusion}\label{s.conc}

We  have investigated a simplified model of a dark matter candidate to gain insight into models where interactions with a top partner are important.  This applies, for example, to stop coannihilation in the MSSM.
 
We showed that for the supersymmetric $\tilde{B}$ coannihilating with the $\tilde{t}_R$, in the low mass region, $m_\chi \lesssim 500$ GeV, channels other than the final state $gg$ can be important and that ignoring these channels can lead to a shift of approximately 10 GeV in the expected mass splitting between $m_\chi$ and $m_{\tilde{t}_R}$.  The LHC has the potential to cover the coannihilation scenario up to $m_\chi\approx$ 500 $\GeV$, and a 100 TeV collider can cover the entire range of dark matter masses consistent with the relic density, $m_\chi \lesssim 2$ TeV. Unfortunately,  the expected rate of dark matter interactions in direct detection experiments remains too low for a discovery of this minimal supersymmetric scenario, even when loop mediated processes are considered. Confirming the dark matter interpretation of a possible collider signal by additional astrophysical observations is going to be very challenging. Conversely, a signal in any direct detection experiment  can only be accommodated within this simple MSSM scenario if additional new physics is at the TeV scale.

Allowing for a free $y_\chi$, one can determine the required coupling to obtain a consistent relic density for any given mass splitting between $\chi$ and $\tilde{t}$ as a function of the dark matter mass, $m_\chi$. We note that $\lambda_h$ is much less powerful than $y_\chi$ for cosmology and only relevant in the small  $r = m_{\tilde{t}} /   m_\chi$, region. However, $\lambda_h$ can have a significant impact on gluon fusion, which is the main production mode for the 125 GeV Higgs boson. Collider searches are insensitive to the couplings $y_{\chi}$ and $\lambda_h$, but depend on the mass splittings, therefore a large region of the parameter space under consideration will be probed by LHC14 and, more comprehensively, at a 100 TeV collider. 

We also computed the loop induced coupling of a pair of $\chi$ to the Higgs boson, which is particularly relevant for the direct detection cross-section. The direct detection cross-section scales with $y_\chi^4$ and depends on $m_\chi$, $r$ and $\lambda_h$.   While current bounds from LUX are not able to constrain this scenario, we find that the sensitivity of near future direct detection experiments, namely LZ and XENON1T, will allow for the testing of this scenario. 

In conclusion, we showed that our Simplified Model has a rich and interesting phenomenology. Current experimental limits leave much of the parameter space untested.
However, it is interesting to note that a combination of 
future collider
 direct searches and indirect and direct dark matter detection experiments will comprehensively 
 probe the
 parameters of this model.

\section*{Acknowledgements}
SV thanks C. Hellmann and  A. Hryczuk for discussions regarding the Sommerfeld effect. AP thanks G.~Giudice for helpful communications.
The work of AI was supported in part by the DFG cluster of excellence ``Origin and Structure of the Universe." This material is based upon work supported by the U.S. Department of Energy, Office of Science, Office of High Energy Physics under Award Number DE-SC0007859.
AP and NS thank the KITP, where this research was supported in part by the National Science Foundation under Grant No. NSF PHY11-25915.  NS thanks the Aspen Center for Physics and the NSF Grant \#1066293 for hospitality during the completion of this work. 

\bibliography{Stops}
\bibliographystyle{utphys}
\end{document}